\RequirePackage{fix-cm}
\documentclass[11pt,a4paper]{article}
\usepackage{graphicx}
\usepackage[reqno]{amsmath}
\usepackage{amssymb}
\usepackage{mathtools}
\usepackage{mathrsfs}

\usepackage[hyphens]{url}
\usepackage{hyperref}

\usepackage{graphicx}
\usepackage{subcaption}

\usepackage{multirow}
\usepackage{makecell}
\usepackage{float}
\usepackage{booktabs}
\usepackage{tabularx}
 
\usepackage{authblk}
\usepackage[todonotes={textsize=tiny}]{changes}
\definechangesauthor[color=yellow]{DH}
\definechangesauthor[color=orange]{SS}
\definechangesauthor[color=green]{LM}
\usepackage{booktabs, makecell, tabularx}
\usepackage{rotating}
\usepackage[export]{adjustbox}

\providecommand{\keywords}[1]
{
  \small	
  \textbf{\textit{Keywords--}} #1
}

\newcommand*\diff{\mathop{}\!\mathrm{d}}

\title{Accurate Distances Measures and Machine Learning of the Texture-Property Relation for Crystallographic Textures Represented by One-Point Statistics} 
\author[1, 2]{Tarek Iraki*}
\author[3]{Lukas Morand}
\author[2]{Norbert Link}
\author[1, 4]{Stefan Sandfeld}
\author[3]{Dirk Helm}

\affil[1]{Institute for Advanced Simulations - Materials Data Science and Informatics (IAS-9), Forschungszentrum Jülich GmbH, Jülich, Germany}
\affil[2]{Intelligent Systems Research Group ISRG, Karlsruhe University of Applied Sciences, Moltkestr. 30, 76133 Karlsruhe, Germany}
\affil[3]{Fraunhofer Institute for Mechanics of Materials IWM, Wöhlerstraße 11, 79108 Freiburg, Germany}
\affil[4]{Faculty of Georesources and Materials Engineering, RWTH Aachen University, Germany}
\affil[*]{corresponding author, email: t.iraki@fz-juelich.de}

\begin{document}

\date{}
\maketitle

\begin{abstract}
The crystallographic texture of metallic materials is a key microstructural feature that is responsible for the anisotropic behavior, e.g., important in forming operations. In materials science, crystallographic texture is commonly described by the orientation distribution function, which is defined as the probability density function of the orientations of the monocrystal grains conforming a polycrystalline material. For representing the orientation distribution function, there are several approaches such as using generalized spherical harmonics, orientation histograms, and pole figure images . Measuring distances between crystallographic textures is essential for any task that requires assessing texture similarities, e.g. to guide forming processes. Therefore, we introduce novel distance measures based on (i) the Earth Movers Distance that takes into account local distance information encoded in histogram-based texture representations and (ii) a distance measure based on pole figure images. For this purpose, we evaluate and compare existing distance measures for selected use-cases. The present study gives insights into advantages and drawbacks of using certain texture representations and distance measures with emphasis on applications in materials design and optimal process control. 

\end{abstract}

\section{Introduction}

\subsection{Motivation} 

The crystallographic texture is a microstructural feature that describes the non-uniform distribution of grain orientations in polycrystalline materials. In metallic materials, the texture is initiated via solidification and can be modified by thermal, mechanical, and termo-mechanical processes like heat treatment, cold forming or hot forming.The existence of preferred grain orientations, however, leads to the typically anisotropic behavior of processed metallic products and is therefore an important aspect for the metal industry. Consequently, texture optimization along the processing-structure-properties chain \cite{olson1997computational} is of major importance. The measurement of the distance or also the similarity between crystallographic textures becomes essential when target textures have to be reached by a process. Then it is important to know how far the actual texture of the process path is away from the target \cite{dornheim2020structureguided}. The knowledge of texture distances is also required for the design of textures for given desired properties \cite{iraki2021multi}. 

In general, the distribution of microstructural features, such as the orientation distribution, can be described by sets of $n$-point correlation functions \cite{torquato2002random}. This work is focused on the first order approximation by the one-point correlation function (probability density function). This approximation is commonly used for many tasks in materials science and industry, see \cite{Niezgoda-microstructure-variance-2011} and \cite{kocks2000texture}. One of the most important one-point correlation functions for many material properties in polycrystalline materials is the orientation distribution function (ODF), see \cite{kocks2000texture}. It is used in this paper to demonstrate the concepts of representations and distance measurements of one-point correlation functions. In the literature, however, only few investigations in ODF distance measures can be found. These are explained in detail in the subsequent section summarizing the related work, where we also present other density distance measures, which have so far not been applied to ODF. 

The commonly used distance measures take only into account the shape difference of the density functions, but do not consider the underlying orientation distances and correlations. This missing information is represented in the proposed new distance measures, which also represent orientation distances or even neighborhood correlations between orientations. The effect of incorporating or neglecting this information by using the various distance measures is discussed and demonstrated in this paper. We have chosen process path optimization and machine learning of microstructure-properties-relation as demonstration use cases.

\subsection{Related work}
Crucial for measuring texture distances is the way how the crystallographic texture, i.e. the ODF, is represented. Having grain orientation data, the ODF can be approximated basically by using different methods, representing the ODF under different aspects: (i) on the basis of a series expansion using generalized spherical harmonics (GSH) as described in Bunge \cite{bunge2013texture} or hyperspherical harmonics \cite{mason-schuh-2008-Hyperspherical-harmonics,mason-schuh-2009-Expressing-Crystallographic-Textures}, (ii) on the basis of orientation histograms \cite{dornheim2020structureguided}, and (iii) by parameterizing characteristic texture components and fibers, see Delannay et al. \cite{delannay1999new} and Kocks et al. \cite{kocks2000texture}. Also experimental pole figures and inverse pole figures \cite{kocks2000texture} can be used to represent the ODF.

As the ODF is a probability density function, the similarity between two ODFs can be expressed by many density-related distance and divergence measures from probability theory textbooks, such as the four most important measures Kullback-Leibler Divergence \cite{Kullback-Leibler-Divergence-1951}, Hellinger distance \cite{Hellinger-Distance-1909}, Bhattacharyya distance \cite{Bhattacharyya-Distance-1946}, and Wasserstein metric \cite{Wasserstein-metric-1969}.

The distance measure between two ODFs $f^{(1)}(g)$ and $f^{(2)}(g)$ commonly used in material science is defined as \cite{adams2001microstructure}:
\begin{equation}
\mathscr{D}_{\mathrm{odf}}(f^{(1)} (g),f^{(2)} (g)) = \int_g \Bigl(f^{(1)}(g) - f^{(2)}(g)\Bigr)^2 \diff g,
\label{eq:global_comparison}
\end{equation}
with $g$ being an orientation in the orientation space $SO(3)$. This distance measure represents the integral squared difference between the amplitudes of the density functions only. Neither the distance nor the neighborhood relations between the corresponding orientations are reflected by this measure.

Within this distance measure, the ODFs can be expressed by their approximations, as mentioned above. In \cite{adams2001microstructure}, approximations of the ODF distance are calculated similiarly to Eq. (\ref{eq:global_comparison}), but on the basis on two-point correlation functions. In our work, we transfer this approach to ODFs, which capture one-point correlation functions. Instead of expanding into a finite GSH series, the density is approximated by principal component analysis (PCA) \cite{Pearson1901-PCA} in \cite{paulson2017reduced} to reduce the representation dimension of the density functions (representing texture representative volume elements) \cite{kalidindi2011microstructure,niezgoda2011understanding}. The distance is then measured by the Euclidean distance between the PCA coefficient vectors (see \cite{niezgoda2013novel}). The PCA coefficients are ordered according to ascending represented variance of the principal components. In texture analysis, higher variance components are more important, but by using the Euclidean distance all components are weighted equally.

A different distance measure is introduced in \cite{sundararaghavan2005synergy} for fcc textures (e.g. goss, brass, copper, etc. \cite{kocks2000texture}) that are described by their typical components in Rodrigues space. \cite{sundararaghavan2005synergy} uses pole-density functions of essential orientation fibers in the fundamental zone for decomposition. The ODF is then described by a feature vector containing the expansion coefficients as entries. The distance between two ODFs is then measured using the $l_2$ norm between their two feature vectors. This approach, however, is only applicable for textures that show such characteristic fibers and is therefore not generalizable.

Alternatively, in \cite{engler1994statistics} and \cite{moreau1994optimization}, the ODF distance is measured on the basis of pole figures. This is done by calculating the sum of absolute or squared distance, respectively, between discretized pole figures of the ODFs. Another distance measure based on pole figures was introduced in \cite{Tarasiuk-1996-Comparison-Of-Crystallographic-Textures}: The two values of the pole figures of two ODFs are taken at the same position of the discretization grid and stored as entries of a two dimensional vector. The resulting set of vectors is then plotted as a point cloud in a 2-d diagram. This scatter plot is then analyzed by performing a linear regression of the point cloud, which would be just a diagonal in case that both ODFs were identical. The regression parameters \textit{offset} and \textit{slope} as well as the \textit{correlation factor} are determined, where the latter is used as a similarity score between the two ODFs to classify them as similar, when it is larger than 0.9. However, this measure ignores the neighborhood relations of the orientations represented in the pole figures. 

The idea of using orientation histograms to measure similarities of textures is used by \cite{EISENLOHR2008670}. This work analyzes the reconstruction of textures by a finite number of orientations with equal weight. In \cite{EISENLOHR2008670}, the similarity between original and reconstructed ODFs are quantified by the measure proposed in \cite{Tarasiuk-1996-Comparison-Of-Crystallographic-Textures}, but also by comparing orientation histograms in the Euler space by the root mean squared deviation of the bin entries of both histograms. The use of histogram bin entries does not consider neighborhood relations between orientations. Furthermore, operating in the Euler space suffers from its distortion.

Dornheim et al. \cite{dornheim2020structureguided} introduced an alternative approach that uses the Chi-Squared distance \cite{Pele2010chi-square-distance} applied to histogram-based texture representations in the space of unit quaternions. By using unit quaternions to describe orientations, distances between individual orientations can be evaluated in a straightforward manner \cite{Huynh2009}. Moreover, the Chi-Squared distance measure has the particular advantage that it is statistically consistent and minimum and maximum bounds exists. But as before, also only bin entries are compared. Neglecting the underlying orientation distances leads to a distance measure, which does not represent texture distances well, especially in the case of sparse histograms. 

A histogram based measure also taking into account the distances between the representatives of the bins is the Earth Mover's Distance (EMD) \cite{Tomasi-1997-EMD-MDS-Image-Retrieval, Rubner-2000-EMD-Image-Retrieval}. Therefore it is better suited to represent texture distances even with sparse histograms. 

\subsection{Contribution} 
The main contribution of the present paper is the development of novel distance measures for crystallographic textures. These are (i) the Earth Movers distance taking into account local distance information encoded in histogram-based texture representations and (ii) a distance measure based on pole figure images. The GSH distance is used for comparison, for which we provide a complete derivation of this distance measure. We compare the distance measures to existing measures and outline their applicability. Furthermore, we show the effect of selected texture representations on the developed distance measures.

Our second use-case demonstrates the suitability of the selected texture representations for carrying property information. In the field of machine learning in materials science, it is essential to develop predictive models that can be used to design and optimize materials with specific properties. For this purpose, we develop neural networks for learning texture-property relations for the investigated texture representations and investigate the results among each others. 

\subsection{Paper structure}
The methods applied in this paper are summarized in the following Section \ref{s:Methods}. This includes a general introduction to the orientation distribution function in Section \ref{ss:methods_odf}, descriptions of the applied texture distance measures and corresponding texture representations in Section \ref{ss:Texture-Representations-and-Distance Measures}, the machine learning models applied for approximating texture-properties-relations that are used for an additional application study in Section \ref{ss:Structure-Property-Mapping}, and the underlying data used to compare and evaluate the above mentioned methods in Section \ref{texture-data-sets-for-evaluation}. Then, in Section \ref{s:Results}, first, we outline advantages and disadvantages of the proposed texture distance measures and investigate the effect of the corresponding texture representations on the distance measures. Second, as an additional study, the effect of texture representations on learning texture-properties-relations are analyzed. The findings of this work are discussed in Section \ref{s:Discussion}. We summarize our work and draw conclusions in Section \ref{s:Summary}. Important symbols for representing crystallographic texture and measuring texture distances that are used throughout this study are summarized in Table \ref{tab:summary-of-notation} and Table \ref{tab:summary-of-functions}.

\section{Methods}
\label{s:Methods}
\subsection{Orientation distribution function}
\label{ss:methods_odf}
Grain orientations can be parameterized in different ways, see \cite{hansen1978tables}. Most frequently used in experimentation are Euler angles \cite{bunge2013texture,pospiech1972parameter}, which describe an orientation as three subsequently executed rotations around defined axes of a Cartesian coordinate system. However, due to singularities, the Euler space is distorted, in general. To overcome this issue, the axis-angle description \cite{hansen1978tables} or Rodrigues vectors \cite{frank1988orientation} can be used, which describe orientations as a rotation around a 3d-vector (the length of the vector yields the rotation). A general drawback of these descriptions, however, remains, which is that for doing calculus, signs and special conventions have to be respected and the amount of permutations that follows from it is relatively high. For this reason, unit quaternions, which are unique and well-defined, have been introduced for the description of orientations \cite{morawiec1989some}.

To describe the crystallographic texture of materials, typically, the orientation distribution function defined by

\begin{equation}
f(g)\mathrm{d}g = \frac{\mathrm{d} V}{V},
\end{equation}
with $g$ being an orientation in $SO(3)$ and $V(g)$ describing the volume in $SO(3)$, is used. Although our study is not limited by certain symmetry conditions, in this work, we assume $f(g)$ underlying cubic symmetry conditions as we focus particularly on steel materials. As a consequence of such symmetry conditions, certain subspaces of SO(3) are equivalent and, thus, SO(3) can be reduced to the so-called fundamental zone, see \cite{hansen1978tables}. In the following, we present commonly used texture descriptions and methods to measure distances between textures.

For expressing an orientation $g$, several representations exist (cf. \cite{hansen1978tables}). For this study, however, two typically used representations are of importance, which are the Euler angle representation following \cite{bunge2013texture} and the representation using unit quaternions, see for example \cite{morawiec1989some} and \cite{frank1988orientation}. The Euler angles $\varphi_1,\phi,\varphi_2$ define subsequent rotations about the axes X, Z, X of a Cartesian coordinate system, while quaternions $q_0,q_i,q_j,q_k$ form a vector in a four-dimensional vector space \cite{morawiec1989some}.

\begin{table}[H]
	\caption{Summary of important symbols for representing crystallographic texture that are used throughout this study.} 
	\centering
	\begin{tabular}{ l || p{8.0cm}}
		\toprule
		Notation & Description   \\ \midrule
		$g$ & Orientation (point in $SO(3)$). \\ \hline
		$f(g)$ & Orientation distribution function. \\ \hline
		$f_{\boldsymbol{\mathrm{b}}}(g)$ & Approximation of the ODF as orientation histogram. \\ \hline
		$f_{\mathrm{pf}}(g)$ & Approximation of the ODF as polefigure. \\ \hline
		$f_{\mathrm{gsh}}(g)$ & Approximation of the ODF on the basis of generalized spherical harmonics. \\ \hline				
		$q_o$ & Quaternion representation.\\ 
		\bottomrule				
	\end{tabular}			
	\label{tab:summary-of-notation}
\end{table}

\begin{table}
	\caption{Summary of important symbols for measuring crystallographic texture distances that are used throughout this study.} 
	\centering
	\begin{tabular}{l || p{8.0cm}}
		\toprule
		Notation & Description   \\ \midrule 
		$\mathscr{D}_{\mathrm{\chi^2}}(f_{\boldsymbol{\mathrm{b}}}^{(1)} (g), f_{\boldsymbol{\mathrm{b}}}^{(2)} (g))$ & Chi-Squared distance between two orientation histograms. \\ \hline
		$\mathscr{D}_{\mathrm{sh}} (f_{\boldsymbol{\mathrm{b}}}^{(1)} (g), f_{\boldsymbol{\mathrm{b}}}^{(2)} (g))$ & Sinkhorn distance between two orientation histograms. \\ \hline
		$\mathscr{D}_{\mathrm{pf}} (f_{\mathrm{pf}}^{(1)} (g), f_{\mathrm{pf}}^{(2)} (g))$ & Distance between two pole figures. \\ \hline
		$\mathscr{D}_{\mathrm{gsh}} (f_{\mathrm{gsh}}^{(1)} (g), f_{\mathrm{gsh}}^{(2)} (g))$ & Distance between two orientation densities expressed as GSH. \\ 
		\bottomrule						
	\end{tabular}			
	\label{tab:summary-of-functions}
\end{table}

\subsection{Texture Representations and Distance Measures}
\label{ss:Texture-Representations-and-Distance Measures}

\subsubsection{Generalized Spherical Harmonic Functions}
The ODF $f(g)$ can be approximated by a series expansion on the basis of generalized spherical harmonics \cite{bunge2013texture}
\begin{equation}
f_{\mathrm{gsh}}(g) = \sum_{l=0}^{\infty} \sum_{m=-l}^{+l} \sum_{n=-l}^{+l} C_{l}^{mn} T_l^{mn}(g),
\label{eq:GSH_expansion}
\end{equation}
with the orthonormal basis functions
\begin{equation}
T_l^{mn}(\varphi_1,\phi,\varphi_2) =  e^{im\varphi_2} P_{l}^{mn} (\cos(\phi)) e^{in\varphi_1},
\label{eq:GSH_expansion_T}
\end{equation}
in which $g$ is expressed as Euler angles and  $P_l^{mn}$ denotes generalizations of the associated Legendre functions. The coefficients of the series expansion $C_l^{mn}$ are often used as descriptors for crystallographic texture, as they encode the shape of the orientation distribution function, see for example \cite{kalidindi2004microstructure,lyon2004gradient} and \cite{proust2006procedures}. In these works, however, symmetries of the ODF are taken into account, yielding a more compact form of Eq. (\ref{eq:GSH_expansion}) as certain functions $T_l^{mn}(g)$ can be transformed or eliminated, depending on the symmetry \cite{bunge2013texture}. 

The distance between two ODFs $f^{(1)} (g)$ and $f^{(2)} (g)$ can be expressed according to Eq. (\ref{eq:global_comparison}) as

\begin{equation}
\begin{aligned}
	\mathscr{D}_{\mathrm{odf}}(f^{(1)} (g),f^{(2)} (g)) & = \frac{1}{8 \pi^2} \int\limits_{0}^{2\pi} \int\limits_{0}^{\pi} \int\limits_{0}^{2\pi} \\   	
      	& \Bigl(f^{(1)}(\varphi_1, \Phi, \varphi_2) - f^{(2)}(\varphi_1, \Phi, \varphi_2)\Bigr)^2 \\ 
       	& d \varphi_1 \sin(\Phi) d \Phi d \varphi_2.
\end{aligned}
\label{eq:gsh-1}
\end{equation}
When the densities $f^{(1)} (g)$ and $f^{(2)} (g)$ are expanded in two series of GSH base functions

\begin{equation}
f_{\mathrm{gsh}}^{(1)}(g) = \sum_{l=0}^{\infty} \sum_{m=-l}^{+l} \sum_{n=-l}^{+l} C_{l}^{mn} e^{im\varphi_2} P_{l}^{mn} (\cos(\Phi)) e^{in\varphi_1}
\label{eq:gsh-2}
\end{equation}

and

\begin{equation}
f_{\mathrm{gsh}}^{(2)}(g) = \sum_{l=0}^{\infty} \sum_{m=-l}^{+l} \sum_{n=-l}^{+l} \tilde{C}_{l}^{mn} e^{im\varphi_2} P_{l}^{mn} (\cos(\Phi)) e^{in\varphi_1},
\label{eq:gsh-3}
\end{equation}
the distance measure of Eq. (\ref{eq:gsh-1}) becomes 

\begin{equation}
\begin{aligned}
\mathscr{D}_{\mathrm{gsh}}(f_{\mathrm{gsh}}^{(1)}(g),f_{\mathrm{gsh}}^{(2)}(g)) & = \frac{1}{8 \pi^2}  \int\limits_{0}^{2\pi} \int\limits_{0}^{\pi} \int\limits_{0}^{2\pi}\\
         & \Biggl( \sum_{l=0}^{\infty} \sum_{m=-l}^{+l} \sum_{n=-l}^{+l} (C_{l}^{mn} - \tilde{C}_{l}^{mn}) e^{im\varphi_2} P_{l}^{mn} (\cos(\Phi)) e^{in\varphi_1} \Biggr)^2\\
         &  d \varphi_1 \sin(\Phi) d \Phi d \varphi_2.
\end{aligned}
\label{eq:gsh-4}
\end{equation}
Considering the orthogonality of $P_l^{mn}$, the distance measure between orientation densities $f_{\mathrm{gsh}}^{(1)}(g)$ and $f_{\mathrm{gsh}}^{(2)}(g)$ reduces to
\begin{equation} 
\mathscr{D}_{\mathrm{gsh}}(f_{\mathrm{gsh}}^{(1)}(g), f_{\mathrm{gsh}}^{(2)}(g)) = \sum_{l=0}^{\infty} \sum_{m=-l}^{+l} \sum_{n=-l}^{+l} \frac{1}{2l + 1} (C_{l}^{mn} - \tilde{C}_{l}^{mn})^2.
\label{eq:gsh-5}
\end{equation}
For the sake of completeness, we included the full derivation of Eq. (\ref{eq:gsh-5}) in Appendix \ref{sec:Appendix-Gsh-Distance}.

\subsubsection{Orientation Histograms}
\label{sss:ohg-texture-description}

Orientation histograms have been introduced in \cite{dornheim2020structureguided} to facilitate formulating statistical distance measures. The basis of the approach is the quaternion distance $\Phi(q_1,q_2)$ between two unit quaternions $q_1$ and $q_2$ as given in \cite{Huynh2009}:
\begin{equation}
\Phi(q_1,q_2) = \min(|q_1-q_2|, |q_1+q_2|).
\label{eq:quat_distance}
\end{equation}

To construct an orientation histogram, a set $O$ of $j$ nearly uniformly distributed orientations $o_j$ in $SO(3)$ (or the fundamental zone, if symmetry conditions hold) is needed. The set $O$ forms the bin centers of the orientation histograms and is derived using the algorithm introduced in \cite{Quey2018}, as is implemented in the software \textit{neper} \cite{quey2011large}. 

For a set of orientations $G$ that describes the crystallographic texture of a material, assignment vectors $\boldsymbol{w}_g$ are calculated for each element $g \in G$ by
\begin{equation}
w_\mathrm{g}^{(j)} = 
\left\{\begin{array}{ll}
\frac{\Phi(g,o_j)}{\sum_{o_i \in N_l} \Phi(g,o_i)}, & \text{if} ~~ o_j \in N_l \\
0, & \text{else} 
\end{array}\right. ,
\end{equation}
with $w_\mathrm{g}^{(j)}$ denoting the $j$th component of $\boldsymbol{w}_g$. Furthermore, $N_l$ is a set of $l$ nearest neighbor orientations of $g$ in terms of the orientation distance Eq. (\ref{eq:quat_distance}). This so-called soft-assignment approach (with soft assignment factor $l$) guarantees that orientations that do not lie perfectly in the bin center are assigned proportionally to the neighbor bins $o_i \in N_l$. Finally, a volume average over the assignment vectors of the individual orientations $\boldsymbol{w}_g$ yields the orientation histogram 
\begin{equation}
f_{\boldsymbol{\mathrm{b}}}(g) = \frac{1}{V}\sum_{g \in G} V(g)\boldsymbol{w}_{g}.
\end{equation}
On the basis of the orientation histograms, texture distances can be defined using any histogram-based distance measure. The Chi-Squared distance and the EMD, as well as a faster implementation of the EMD, called the Sinkhorn distance \cite{Cuturi-2013-sinkhorn-distance}, are introduced in the following.

\paragraph{Chi-Squared distance:} In \cite{dornheim2020structureguided}, the Chi-Squared distance measure \cite{Pele2010chi-square-distance} was shown to be suitable for this purpose.  The Chi-Squared distance between two orientation histograms $f_{\boldsymbol{\mathrm{b}}}^{(1)} (g)$ and $f_{\boldsymbol{\mathrm{b}}}^{(2)} (g)$ is defined as follows \cite{Pele2010chi-square-distance}: 
\begin{equation}
\mathscr{D}_{\mathrm{\chi^2}}(f_{\boldsymbol{\mathrm{b}}}^{(1)} (g), f_{\boldsymbol{\mathrm{b}}}^{(2)} (g))) = \sum_{j=1}^J \frac{(f_{\mathrm{b}_j}^{(1)} (g) - f_{\mathrm{b}_j}^{(2)} (g))^2}{f_{\mathrm{b}_j}^{(1)} (g) + f_{\mathrm{b}_j}^{(2)} (g)}.
\label{eq:chi2_whd}
\end{equation}
However, for comparing sparsely populated histograms, the applicability of the Chi-Squared distance measure is limited. In such cases, the Chi-Squared distance measure is very likely to show very high values (close to the maximum) and is therefore not sensitive for changes in the histograms. This is why we propose to use the EMD \cite{Tomasi-1997-EMD-MDS-Image-Retrieval}. 

\paragraph{Earth movers distance:} The EMD \cite{Tomasi-1997-EMD-MDS-Image-Retrieval, Rubner-2000-EMD-Image-Retrieval}, going back to the works \cite{werman1985distance} and \cite{peleg1989unified}, is based on a solution to the well-known transportation problem \cite{Hitchcock-1941-transportation-problem}. It is typically used for comparing distributions over a discrete region (e.g. the bins of a histogram as being used in image retrieval \cite{Rubner-2000-EMD-Image-Retrieval}). EMD measures the minimum amount of work needed to transport the probability mass from one distribution to the other to make them equal. The work of transfer between two bins is the transported probability mass times the distance between the representatives of the bins. In our case we consider the minimum amount of work needed to transform one orientation histogram into the other. Thereby, orientation distances between bins play an important role, compared to the Chi-Squared distance. 

The computation of the EMD can be formalized as a linear programming problem with constraint conditions as described in \cite{Tomasi-1997-EMD-MDS-Image-Retrieval}, which is used in our approach. The principle procedure of this approach is as follows: Given two sets of tuples: $\boldsymbol{t}^{(1)} = \{ (q^{(1)}_1, f_{\mathrm{b}_1}^{(1)} (g)), (q^{(1)}_2, f_{\mathrm{b}_2}^{(1)} (g)), ..., (q^{(1)}_m, f_{\mathrm{b}_m}^{(1)} (g)) \}$ and $\boldsymbol{t}^{(2)} = \{ (q^{(2)}_1, f_{\mathrm{b}_1}^{(2)} (g)), (q^{(2)}_2, f_{\mathrm{b}_2}^{(2)} (g)), ..., (q^{(2)}_n, f_{\mathrm{b}_n}^{(2)} (g))\}$ where $q^{(1)}_i$, $q^{(2)}_j$ are the histogram bin centers represented as unit quaternions and $f_{\mathrm{b}_i}^{(1)}(g)$, $f_{\mathrm{b}_j}^{(2)} (g)$ are the entries (weights) of the corresponding histogram. Respecting the crystal symmetry, the ground distance matrix $D=[d_{ij}]$ is calculated, where $d_{ij}$ is the distance between the quaternion representations $q^{(1)}_i$ and $q^{(2)}_j$. The quaternion distance \cite{Huynh2009} is used as the distance function: 

\begin{equation}
d_{ij} = \arccos(|q^{(1)}_i \cdot q^{(2)}_j|),
\label{eq:quat-distance-emd}
\end{equation}
where $\cdot$ denotes the product of the vector elements (not the quaternion multiplication, which results in another quaternion). The aim of solving the transportation problem is to find a flow matrix $F=[f_{ij}]$, where $f_{ij}$ is the flow between $f_{\mathrm{b}_i}^{(1)} (g)$ and $f_{\mathrm{b}_j}^{(2)} (g)$, that minimizes the overall work 

\begin{equation}
WORK (\boldsymbol{t}^{(1)}, \boldsymbol{t}^{(2)}, F) = \sum_{i=1}^m \sum_{j=1}^n f_{ij} d_{ij}.
\label{eq:emd-minimze-1}
\end{equation}
Once the transportation problem is solved, and thus the optimal flow $F$ is found, the EMD is defined as the normalized distance between the sets $\boldsymbol{t}^{(1)}$ and $\boldsymbol{t}^{(2)}$:
\begin{equation}
\mathscr{D}_{\mathrm{emd}} (\boldsymbol{t}^{(1)}, \boldsymbol{t}^{(2)}) = \frac{\sum_{i=1}^m \sum_{j=1}^n f_{ij} d_{ij} } {\sum_{i=1}^m \sum_{j=1}^n f_{ij}}.
\label{eq:emd-minimze-2}
\end{equation}

\paragraph{Sinkhorn distance:} The solution of the transport problem as discussed above suffers from the increase in complexity with increasing number of histogram bins, which is also the dimensionality $\delta$ of the histogram representation space. If two histograms of dimension $\delta$ are compared, the costs for the calculation of the optimal transport distances scale at least in $O({\delta}^3 log(\delta))$ \cite{Cuturi-2013-sinkhorn-distance}, \cite{pele-2009-emd}. The transportation problem therefore becomes intractable when $\delta$ exceeds around 100. To overcome this issue, in \cite{Cuturi-2013-sinkhorn-distance} the Sinkhorn algorithm is proposed instead. In this work, an additional constrain is introduced, in order to ensure that the the KL divergence between the flow matrix $F$ and $\pi_i$ $\zeta_j^T$ is smaller to the pre-defined parameter $\alpha$:

\begin{equation}
KL(F, \pi_i \zeta_j ^T ) \leq \alpha \quad \textrm{with} \quad \zeta_j = \sum_i f_{i,j} \quad \textrm{and} \quad \pi_i = \sum_j f_{i,j}.
\label{eq:kl-1}
\end{equation}
This results in the requirement that $s(F)$ should be as large as possible, 

\begin{equation}
s(F) = \sum_{i,j} f_{i,j} \log f_{i,j} \Rightarrow MAX.
\label{eq:kl-2}
\end{equation}
The resulting (dual) Sinkhorn distance is
\begin{equation}
\mathscr{D}_{\mathrm{sh}} = \min (\sum_{i,j} f_{i,j} \cdot d_{i,j} - \frac{1}{\lambda} s(F)),
\label{eq:sinkhorn}
\end{equation}
which is again a convex function. By forming the Lagrangian and searching for its minimum with respect to $f_{i,j}$, a strictly convex problem is obtained yielding a unique optimal solution. It is shown that the resulting optimum is also a distance. Since the Sinkhorn algorithm relies on matrix-vector products, the distance can be computed through Sinkhorn’s matrix scaling algorithm, whereby it is several orders of magnitude faster than transport solvers. 

In our approach, we calculate the histogram distances using the highly efficient Sinkhorn algorithm (\cite{Cuturi-2013-sinkhorn-distance}, Algorithm $1$) with $100$ iterations and setting the value of parameter $\lambda$ to $0.01$. Furthermore the algorithm can be integrated as a fully differentiable function in optimization frameworks such as being used for neural networks and implemented on Graphics Processing Units.

\subsubsection{Pole figures}
\label{sss:pole-figures}
A pole figure is a graphical representation used in materials science and crystallography to display the orientation of polycrystals. It is based on a stereographic projection that maps three-dimensional information on a two-dimensional plane \cite{kocks2000texture}. When using pole figures, the neighborhood relations of the orientations are represented by the neighboring pixels in the image. In this study, pole figures are generated using the software \textit{mtex} \cite{Bachmann-Mtex-2010}. Distances between textures are measured by comparing two pole figures $f_{\mathrm{pf}}^{(1)} (g)$ and $f_{\mathrm{pf}}^{(2)} (g)$ pixel-wise by means of the absolute or $l_1$ loss:

\begin{equation}\label{eq:polefigure-distance} 
	\mathscr{D}_{\mathrm{pf}} (f_{\mathrm{pf}}^{(1)} (g), f_{\mathrm{pf}}^{(2)} (g)) = \frac{1}{M} \frac{1}{N} \sum_{m=1}^M \sum_{i=1}^N | f_{\mathrm{pf}_{m,i}}^{(1)} (g) - f_{\mathrm{pf}_{m,i}}^{(2)} (g) |, 
\end{equation} 
where N denotes the number of pixels and M denotes the number of pole figures.

\subsection{Texture representations for learning texture-property relations}
\label{ss:Structure-Property-Mapping}
When texture-property relations are learned, the textures are transformed into a latent space, which arranges the texture representations according to their properties. Looking at machine learning in materials design, we want to retrieve texture, which correspond to given properties. For this purpose it is very important that the latent space representation also reflects the distance in the original texture space. This can be enforced by using e.g. siamese neural networks \cite{iraki2021multi}. It still remains crucial that the original texture representation contains the essential information about the properties. 

\begin{figure}[H]
  \centering
  \includegraphics[width=0.99\textwidth]{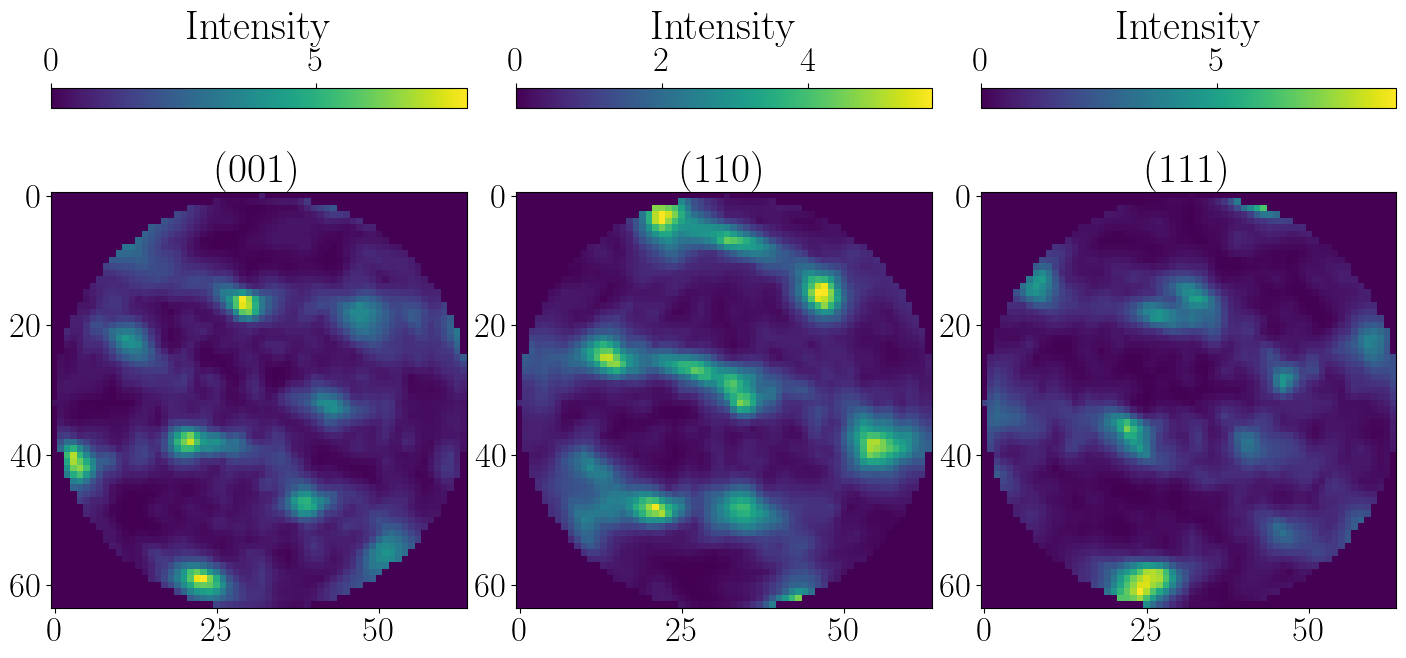}
  \caption{Representation of the ODF by two-dimensional pole figures for the Miller's indices (001) (left), (110) (center) and (111) (right) with a resolution of $64 \times 64$ pixels.} 
  \label{img:pole-figures}
\end{figure}

In order to investigate the ability of the discussed texture representations being suitable for carrying property information, we develop neural network-based models for learning the relation between textures and their corresponding properties, based on the \textit{texture-property data set} described in \ref{ss:Texture-data-generation}. This forward mapping from textures to properties is modeled as a regression problem. The regression loss is given by the mean squared error between the predicted properties $\boldsymbol{\hat{p}}$ and the true properties $\boldsymbol{p}$:

\begin{equation}\label{eq:regr-loss} 
	\mathscr{L}_{\mathrm{regr}} (\boldsymbol{p}, \boldsymbol{\hat{p}}) = \frac{1}{N} \sum_{i=1}^N ({p_i} - {\hat{p}_i} )^2 + \lambda R(\boldsymbol{\theta}),
\end{equation} 
where $\boldsymbol{\theta}$ are the trainable parameters and $R(\boldsymbol{\theta})$ is a regularization term that is used to prevent overfitting with the hyperparameter $\lambda$ defining the strength of the regularization (also known as weight decay, see \cite{Krogh1991Weight-Decay} and \cite{Hinton1987Weight-Decay}). We use E$_{11}$, E$_{22}$, E$_{33}$, R$^*_{11}$, R$^*_{22}$, R$^*_{33}$ to describe the elastic and plastic of material properties. For learning the relation between textures and corresponding properties, we represent the textures in three different ways:

\begin{itemize}
\item as orientation histograms having the resolution $J=512$ and $l$ $\in $ \{$1$, $3$\}.
\item as pole figures for the Miller's indices $(001)$, $(110)$, $(111)$, where the polar coordinates are transformed into Cartesian coordinates with a resolution of $64 \times 64$ pixels. The resulting pole figure representation for one exemplary ODF is depicted in Fig. \ref{img:pole-figures}.
\item as GSH textures for degree $L$=16 and $L$=10. For the mapping, we use only a subset of the GSH coefficients, as the first coefficient equals one and half of the remaining coefficients are complex conjugates of the other half \cite{tallman2019gaussian}.
\end{itemize}

Fully connected neural networks are used for orientation histograms and GSH representations, whereas convolutional neural networks (CNN) are used for pole figure representations \cite{Yamanaka-2020-polefigures-deep-learning, Koenuma-2020-PF-DL}. CNNs are well-studied deep learning methods for image data \cite{LeCun-1989-CNN, LeCun-1998-CNN,LeCun-1999-CNN} and are particularly powerful for processing images, where the underlying neighborhood relations between pixels are exploited, as is the case with pole figures. The topologies of the neural network models are listed in Tab. \ref{tab:Spm-Topologies}, visualized in Fig. \ref{img:Spm-Topologies} and are explained in more detail in the following:

\paragraph{Orientations histograms:} The topology for the neural network on the basis of orientation histogram (see Tab. \ref{tab:Spm-Topologies} (a) and Fig. \ref{img:Spm-Topologies} (a)) consists of three hidden fully connected layers, where each of the layers halve the size of the previous layer. The output of the third hidden layer is followed by two processing layers which feed a final linear regression layer, giving a float output vector for the six predicted properties $\boldsymbol{\hat{p}}$. 

\paragraph{GSH:} The topology for GSH with $L$=10 (see Tab. \ref{tab:Spm-Topologies} (b) and Fig. \ref{img:Spm-Topologies} (b)) half the size of the previous layer within the first three fully connected hidden layers. This is followed by the final linear regression layer for the six predicted properties $\boldsymbol{\hat{p}}$. The first fully connected hidden layer of the topology for GSH with $L$=16 (see Tab. \ref{tab:Spm-Topologies} (c) and Fig. \ref{img:Spm-Topologies} (c)) reduces the input data to the size of $160$ neurons. This is followed by four hidden fully connected layers, where each of the layers halve the size of the previous layer. This is followed by the final linear regression layer for the six predicted properties $\boldsymbol{\hat{p}}$.

\paragraph{Pole figures:} The layout of the CNN topology can be seen in Tab. \ref{tab:Spm-Topologies} (d) and Fig. \ref{img:Spm-Topologies} (d). The convolutional layers have a filter kernel size of $3 \times 3$, for which an increasing size of the feature maps was selected. A common approach is a stride of length one with no sub-sampling in the convolutional layers. Instead, sub-sampling is performed in the pooling layers with stride of length two, halving the pooling layer input (in a Gaussian resolution pyramid). The convolutional layers and the pooling layers are stacked two times and their output is vectorized. This high-dimensional vector is processed by four consecutive fully connected layers to feeding the final linear regression layer for the six predicted properties $\boldsymbol{\hat{p}}$.

\paragraph{Implementation details:} In the context of hyperparameter optimization the random search method \cite{Bergstra2012RandomSearch} is applied using 5-fold cross-validation. Based on the analysis of the hyperparameters, the following decisions are made about the investigated hyperparameters: The Xavier method \cite{Glorot2010WeightInit} is used for weight initialization. The Adam Optimizer \cite{Kingma2015Adam} was found to be most efficient, with the following parameters: learning rate = $0.001$, weight decay = $10^{-6}$, batch size = $32$. The models are trained for 200 epochs, where the best intermediate result of the test set is retained to prevent over fitting. This can be considered as an application of early stopping \cite{Prechelt2012EarlyStopping}. The $tanh$ activation function is used. The neural networks are implemented based on the Pytorch API \cite{Pytorch-2019}.

\subsection{Texture data sets for evaluation}
\label{texture-data-sets-for-evaluation}

\subsubsection{Texture evolution and properties calculation}
\label{sec:materialmodel}

The basis for the investigations in the present paper is a simulated metal forming process, which is described in \cite{dornheim2020structureguided} and for which a simulation framework has been published on the Fraunhofer Fordatis repository \cite{sim2021}. The simulation is based on the Taylor-type crystal plasticity model introduced in \cite{Kalidindi.1992}, which is modified in terms of hardening as described in \cite{baiker2014determination}. For a detailed description of the material model, we refer to Appendix \ref{sec:Appendix-Material-Model}.

Basically, the simulation framework allows for applying sequences of tension and compression operations on a point within a workpiece of the  material under consideration (which we call material point in the following). The operations are defined by the deformation gradient expressed as a matrix with orthogonal basis vectors $\boldsymbol{e}_i$ 
\begin{equation}
\label{eq:tension_op}
\boldsymbol{F} = 
\left[ \begin{array}{ccc}
F_{11} \boldsymbol{e}_1 \otimes \boldsymbol{e}_1 & 0 & 0 \\
0 & F_{22} \boldsymbol{e}_2 \otimes \boldsymbol{e}_2 & 0 \\
0 & 0 & F_{33} \boldsymbol{e}_3 \otimes \boldsymbol{e}_3 \\
\end{array} \right],
\end{equation}
in which $F_{11}$ is defined by the applied tensile load increment and both, $F_{22}$ and $F_{33}$, are determined such that the stress boundary conditions are fulfilled, cf. \cite{dornheim2020structureguided}. 
Each of the individual tensile load increments $\boldsymbol{F}$ can be oriented arbitrarily to the material point yielding a rotated form of $\widetilde{\boldsymbol{F}}$ following
\begin{equation}
\label{eq:tension_op_rot}
\widetilde{\boldsymbol{F}} = \boldsymbol{R}\boldsymbol{F}\boldsymbol{R}^\top,
\end{equation}
with the rotation matrix $\boldsymbol{R}$. We use the framework to simulate texture evolution and generate different crystallographic textures for our analysis. It is well known that Taylor-type  mean-field models typically overestimate texture evolution \cite{kocks2000texture}. However, for the purpose of our study the used mean-field model is sufficient as we prioritize execution speed over predicting highly realistic material behavior.

In addition, the simulation framework is used to calculate material properties. The properties we focus on are the orientation dependent Young's moduli and anisotropy similar to R-values, as in \cite{morand2022efficient}. R-values, however, are originally developed to express plastic anisotropy in sheet metals. On this basis, we use an R-values-like measure, $\tilde{R}$, to described the plastic anisotropy at the material point in three space directions. To do so, tension tests are conducted in each space direction using the simulation framework. The $\tilde{R}_i$-values are given by the relation between tensile and lateral strain. On the basis of the simulated tension tests, also, the Young's modulus $E_i$ is calculated in three space directions. 

\subsubsection{Crystallographic texture data generation}
\label{ss:Texture-data-generation}
In total, two data sets are generated: a first one, which serves to compare the different measures of texture distances and a second (larger) one to serve the demonstration use case of learning texture-property relations. The generation of both data sets is described in the following. 

\paragraph{Distance measurement evaluation set:}
For this purpose we are seeking datasets where the texture distance between subsequent elements is continuously increasing. To generate such an ODF dataset, we simulate an uniaxial tension test (20 load steps with $4 \%$ applied strain).  
When repeatedly applying tension operations on a material with initial texture as in the example process, the distance from the initial texture is expected to be smoothly and monotonically increasing. We generate three such datasets by starting the simulated tension test with different initial texture as follows:
\begin{itemize}

\item Initial Texture 1: A grey initial texture is used that consists of 512 nearly uniformly distributed orientations ($n_\mathrm{oris}=512$) in the cubic-triclinic fundamental zone. The orientations are generated using the algorithm described in \cite{quey2018nearly}.
\item Initial Texture 2: 100 orientations ($n_\mathrm{oris}=100$) are randomly sampled from the $512$ orientations of Initial Texture 1.
\item Initial Texture 3: The third initial texture consists of only one orientation ($n_\mathrm{oris}=1$).
\end{itemize}

\paragraph{Texture-property data set:} For learning of a generalized texture-property mapping, a total of 76980 simulations were conducted, each by applying seven tension steps of $10 \%$ strain based on Equation \ref{eq:tension_op_rot}. The rotation matrix $\boldsymbol{R}$ is determined on the basis of one randomly chosen orientation from a set of 25 (also nearly uniformly) distributed orientations. The properties $E_i$ and $\tilde{R}_i$ are then calculated for each texture of the thereby obtained texture dataset.

\section{Results}
\label{s:Results}
In the following, we show the effect of using different texture representations in the two investigated use cases.

\subsection{Comparing measures of texture distances}
\label{ss:Texture-Distance_Measurement}
The effect of the various discussed ODF representations on the corresponding ODF distance measures are shown in the following. For this purpose, we use the \textit{distance measurement evaluation set} $\{f^{(0)}(g), f^{(1)}(g), \dots , f^{(19)}(g)\}$ described in \ref{ss:Texture-data-generation}, resulting from the forming process, which consists of $20$ tension steps. We determine the distances from the initial texture $f^{(0)}(g)$ to each subsequent texture $f^{(i)}(g), i=1, \dots , 19$: $\mathscr{D}(f^{(0)}(g), f^{(1)}(g))$, \dots , $\mathscr{D}(f^{(0)} (g), f^{(19)} (g))$, where $\mathscr{D}$ represents the distance function. Such sequences of distance measure values are then compared using different texture representations.

\subsubsection{Orientation histogram and pole figure representations}
\label{sss:Texture-Distance-Measurement-Orientation-histograms-and-Polefigures}
\begin{figure}[H]
	\centering
	\includegraphics[width=0.99\textwidth]{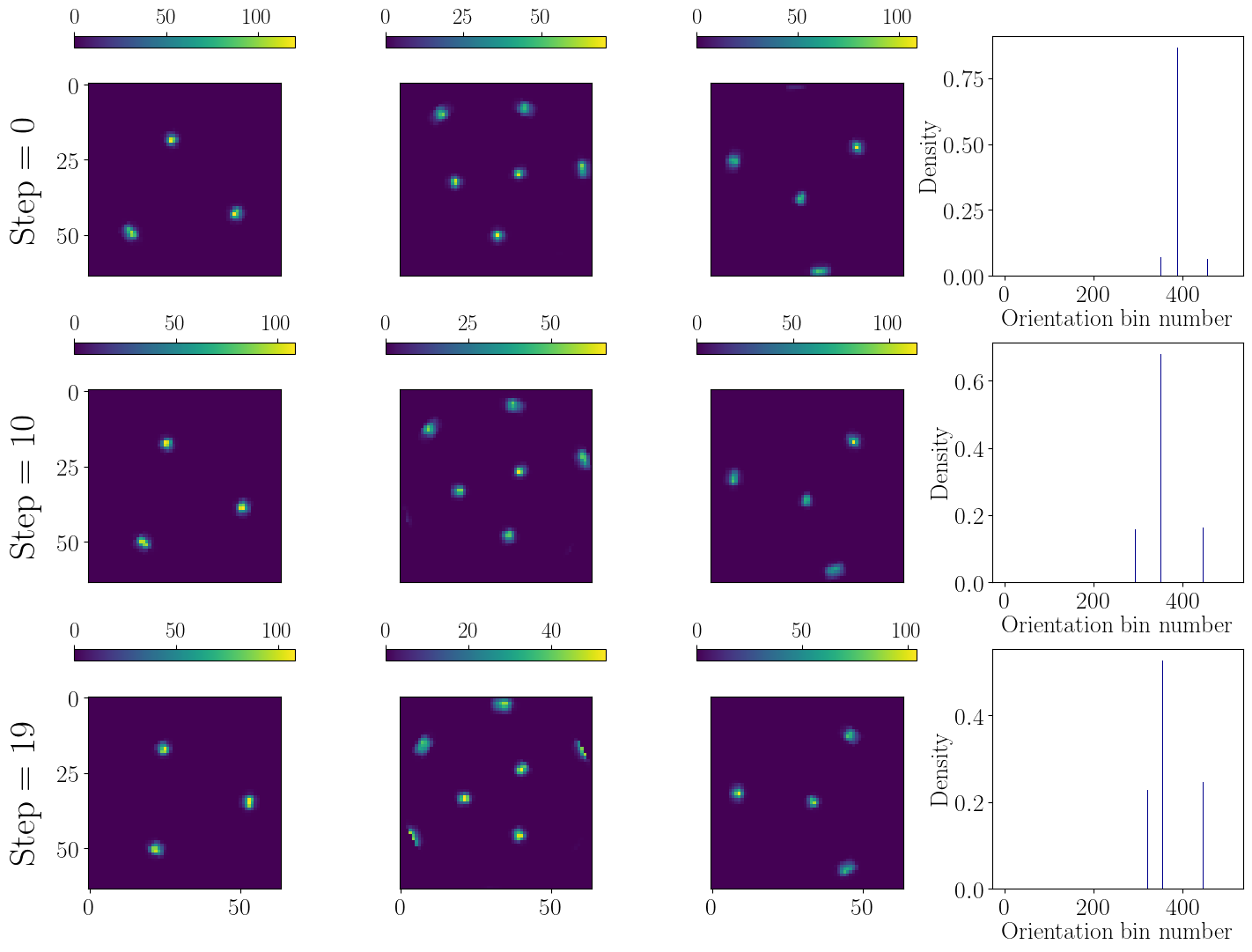}          
  	\caption{Representation of the ODF with one single orientation ($n_{\mathrm{oris}}=1$) by two-dimensional pole figures for the Miller's indices (001), (110), and (111) with a resolution of 64 $\times$ 64 pixels as well as by orientation histograms for number of bins $J=512$ and soft-assignment $l=3$. The first row shows the initial texture at process step $0$, the second row shows the representations at process step $10$, and the third row shows the representations at the final process step $19$.}
  \label{img:texture-examples-nbase-1}
\end{figure}

\begin{figure}[H]
 	\centering
	\includegraphics[width=0.99\textwidth]{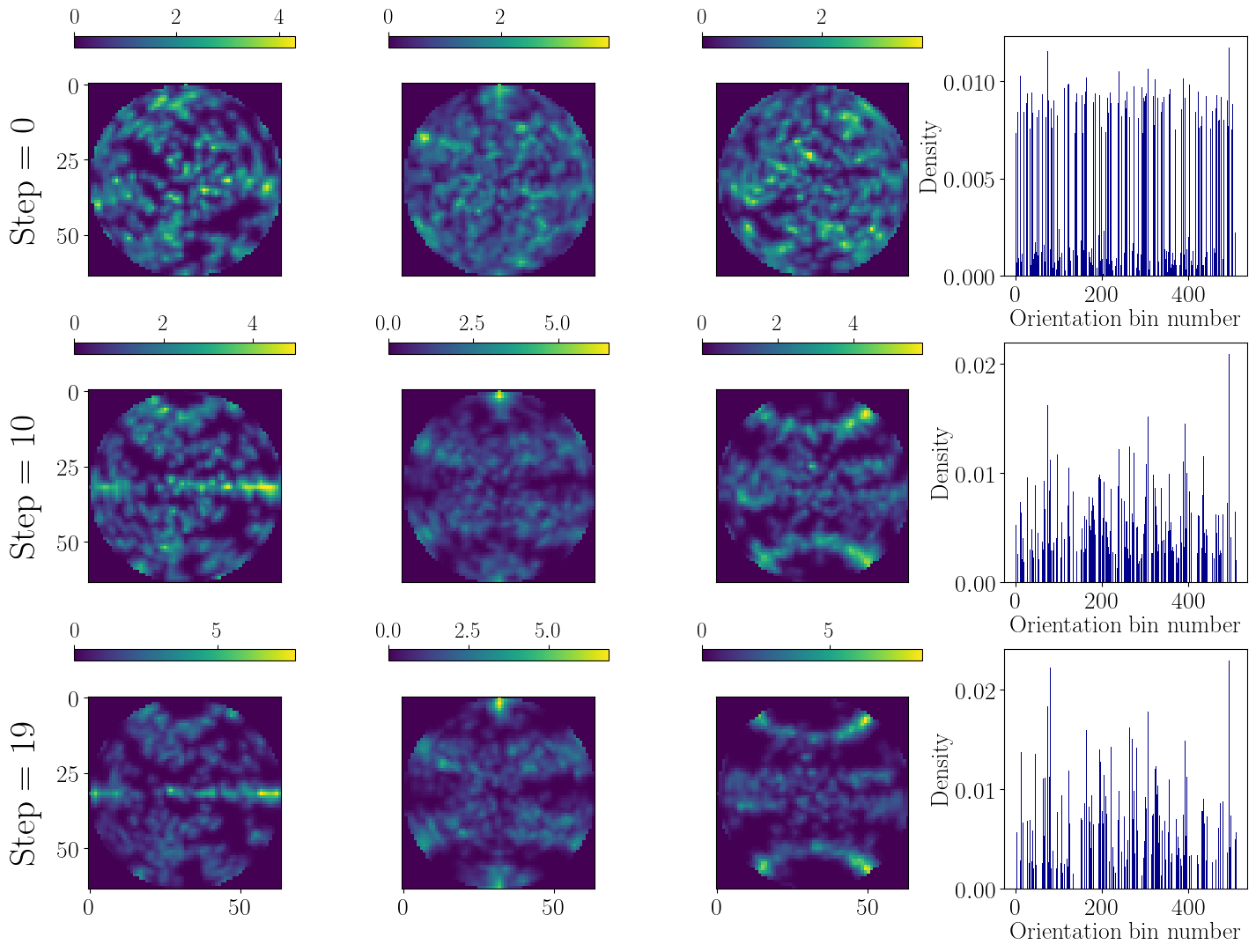}    
	\caption{Representation of the ODF with 100 orientations ($n_{\mathrm{oris}}=100$) by two-dimensional pole figures for the Miller's indices (001), (110), and (111) with a resolution of 64 $\times$ 64 pixels as well as by orientation histograms for number of bins $J=512$ and soft-assignment $l=3$. The first row shows the initial texture at process step $0$, the second row shows the representations at process step $10$, and the third row shows the representations at the final process step $19$.}
  \label{img:texture-examples-nbase-100}
\end{figure}

\begin{figure}[H]
     \centering
	\centering
	\includegraphics[width=0.99\textwidth]{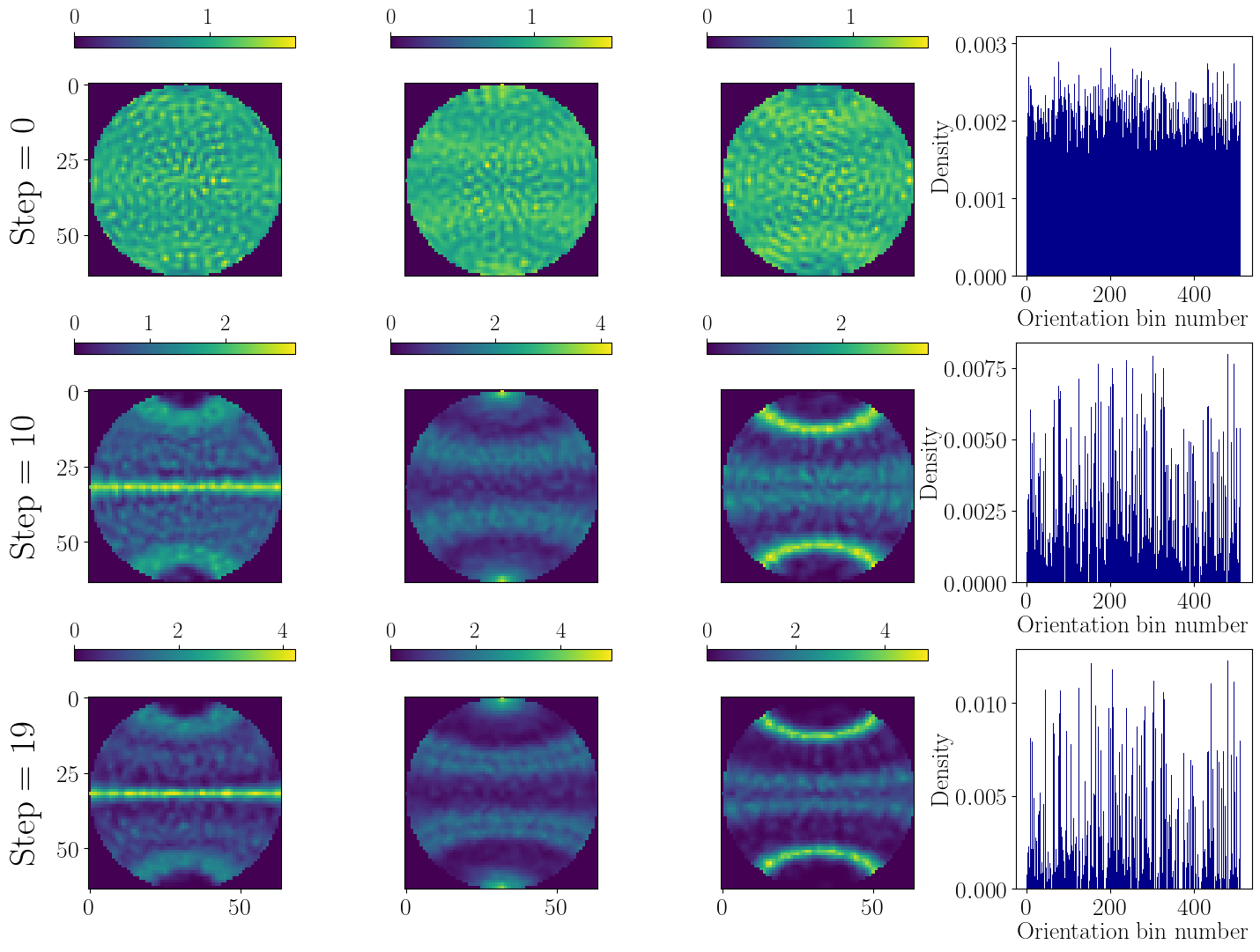}    
         
  \caption{Representation of the ODF with 512 orientations ($n_{\mathrm{oris}}=512$) by two-dimensional pole figures for the Miller's indices (001), (110), and (111) with a resolution of 64 $\times$ 64 pixels as well as by orientation histograms for number of bins $J=512$ and soft-assignment $l=3$. The first row shows the initial texture at process step $0$, the second row shows the representations at process step $10$, and the third row shows the representations at the final process step $19$.}
  \label{img:texture-examples-nbase-512}
\end{figure}

In the following, we compare the distances, which result from using 
\begin{enumerate}
\item the Chi-Squared distance $\mathscr{D}_{\mathrm{\chi^2}}$ (Eq. (\ref{eq:chi2_whd})), 
\item the Sinkhorn distance $\mathscr{D}_{\mathrm{sh}}$ (Eq. (\ref{eq:emd-minimze-2})) and   
\item the pole figure distance $\mathscr{D}_{\mathrm{pf}}$ (Eq. (\ref{eq:polefigure-distance})). 
\end{enumerate}
These measures are calculated for the textures resulting from the forming process starting at the three different initial textures. For each texture set, the orientation histograms and pole figures (depicted in Fig. \ref{img:texture-examples-nbase-1}, Fig. \ref{img:texture-examples-nbase-100}, and Fig. \ref{img:texture-examples-nbase-512}) are created as follows: The textures are represented by orientation histograms of different resolution, where the number of histogram bins $J$ $\in$ \{$128, 256, 512, 1024, 2048$\}. The histogram of each resolution is smoothed with three different soft-assignment values $l$ $\in $ \{$1, 3, 5$\}. We generate pole figures for the Miller's indices $(001), (110), (111)$ and transform the polar coordinates into Cartesian coordinates with a resolution of $64 \times 64$ pixels.

\begin{figure}[H]
	\centering
	\includegraphics[width=0.99\textwidth]{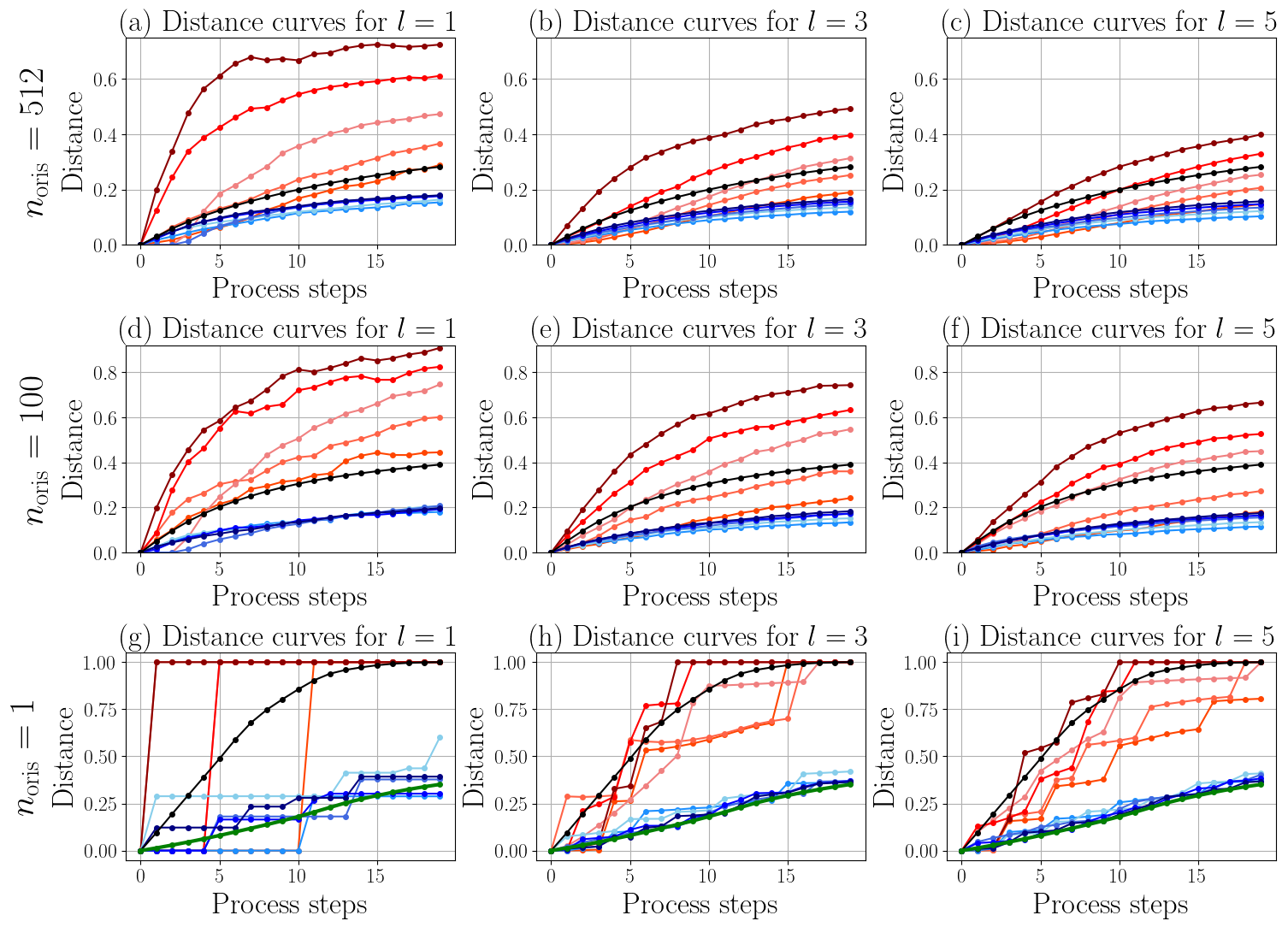}
	\includegraphics[width=0.90\textwidth]{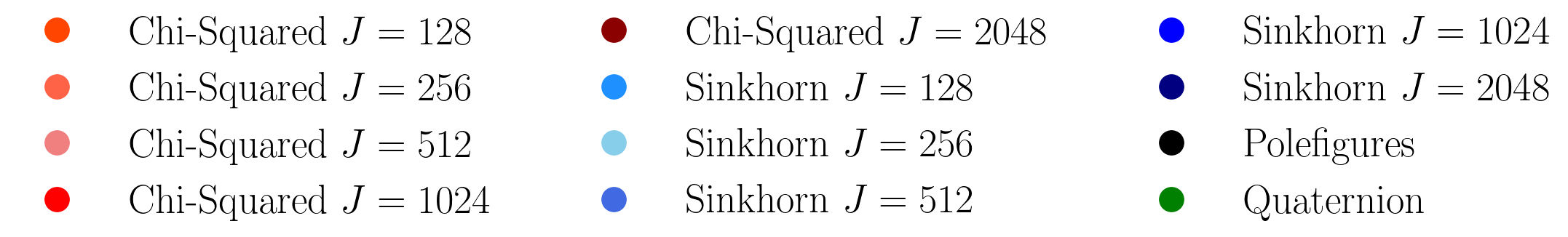}
	\caption{Distances over process step curves for the textures with $512$ orientations ($n_{\mathrm{oris}}=512$, first row), with $100$ orientations ($n_{\mathrm{oris}}=100$, second row) and one single orientation ($n_{\mathrm{oris}}=1$, third row). The Chi-Squared distance (red) and Sinkhorn distance (blue) are represented for the soft-assignment $l=1$ (a, d, g), $l=3$ (b, e, h) and $l=5$ (c, f, i). Each of these diagrams shows the curves for the different number of bins $J \in$ \{$128, 256, 512, 1024, 2048$\} and the pole figure distance curve in black. The Quaternion distance curve is shown in the single orientations diagrams (g, h, i) in green color.}
     \label{img:n-base-512}
\end{figure}

To make the distances comparable independent of the measure under consideration, we normalize the distances by dividing the raw distance by the maximum possible distance in each case. The maximum possible distance is the distance between two single crystals $f_{\mathrm{sc}}^{(0)}(g)$ and $f_{\mathrm{sc}}^{(1)}(g)$, which have the maximum distance in their orientations:

\begin{equation}
\label{eq:max-distance-ohg-polefigure} 
\mathscr{D} = \mathscr{D}(f^{(0)}(g), f^{(k)}(g)) / \mathscr{D}(f_{\mathrm{sc}}^{(0)}(g), f_{\mathrm{sc}}^{(1)}(g)).
\end{equation}
The sequences of these normalized distances are calculated for our three different texture sets and plotted as distance curves over process step number (depicted in Fig. \ref{img:n-base-512}) They are discussed in the following.

\paragraph{Initial Texture 1 ($n_{\mathrm{oris}}=512$):} The distance curves of the Chi-Squared and Sinkhorn distance measures for the ($J$, $l$)-settings of the orientation histograms as well the distance curves of the $l_1$-distance measure for the pole figures are depicted in Fig. \ref{img:n-base-512} (a) - (c). It can be seen that the Chi-Squared and Sinkhorn distance values for $J=512$ and $l=1$ do not increase until step $2$ despite the expected texture evolution due to increasing load (Fig. \ref{img:n-base-512} (a)). This is because in this special case, all texture orientations are located initially in the bin centers. Our process then needs two steps to force the movement of the orientations into another bins. On the other hand, when applying smoothing by soft-assignment with $l=3$ and $l=5$, a small process-enforced movement from the bin center is enough to change the histogram and the distance value. Furthermore, the Chi-Squared distance curve is not smooth for $l=1$ (Fig. \ref{img:n-base-512} (a)) and erroneously not monotonic for $l=1$ and $J=2048$. At higher soft-assignment values $l=3$ (Fig. \ref{img:n-base-512} (b)) and $l=5$ (Fig. \ref{img:n-base-512} (c)) the Chi-Squared distance curve shows a monotonic behavior and is smoother. Another effect of the discretization parameters of the histograms on their distance curves is the variance of the distance values at a process step. One can see from Fig. \ref{img:n-base-512} (a) - (c) that this variance is much higher for the Chi-Squared than for the Sinkhorn distance measure and therefore the Chi-Squared measure is much more sensitive against the choice of histogram parameter values. The Sinkhorn distance as well as the pole figure distance result in monotonic and smoother curves, while the Sinkhorn distance is less sensitive to the number of bins $J$ and soft-assignment $l$. 

\paragraph{Initial Texture 2 ($n_{\mathrm{oris}}=100$):} The results of the investigated distances for the orientation histograms and pole figures are depicted in Fig. \ref{img:n-base-512} (d) - (f). The general behavior of the curves is similar to the case $n_{\mathrm{oris}}=512$, but the curves are rougher for the Chi-Squared measure where also the maximum value is much higher compared to $n_{\mathrm{oris}}=512$ than for the Sinkhorn distance, the curves of which still remain smooth. The distance curve resulting for the pole figure measure remains almost unaffected.  

\paragraph{Initial Texture 3 ($n_{\mathrm{oris}}=1$):} The results of the investigated distances for textures with only one orientation are shown in Fig. \ref{img:n-base-512} (g) - (i). The simulated process delivers a sequence of single crystals where the texture is determined by their orientation quaternions. The quaternion distance Eq. (\ref{eq:quat-distance-emd}) can be used in this case as the ground truth for the texture distance. The quaternion distance curve is strictly monotonic and almost linear. The Chi-Squared distance measure can not cope with this case because it changes its value only if the single crystal orientation moves into a different bin and then immediately takes on the maximum possible value (saturation). Higher-order soft-assignments smoothen this step function behavior and let the curves saturate slower. The discontinuity problem of the Chi-Squared curves is less prominent for Sinkhorn distance curves because the quaternion distances between bin centers are also taken into account. This also forces the Sinkhorn distance curves to be closer to distance curve and also prevents saturation. The pole figure distance curve is monotonic and smooth, but deviates from the linear ground truth curve by forming a half sigmoid function which saturates smoothly with increasing step number. 

\subsubsection{Generalized Spherical Harmonics}
\label{sss:Texture-Distance-Measurement-GSH}
After having learned that the Sinkhorn distance delivers the best results related to histogram distance measures at textures composed of 512 orientations, we now compare these results to the results from using the distance measure $\mathscr{D}_{\mathrm{gsh}}$ (Eq. (\ref{eq:gsh-5}))  based on GSH texture representations of various GSH degrees. For this part of our study, we use the Initial Texture 1 set ($n_{\mathrm{oris}}=512$) from the \textit{distance measurement evaluation set} described in \ref{ss:Texture-data-generation}. The ODF is approximated by a GSH series expansion (\cite{bunge2013texture}) of the degrees L $\in$ \{$4$, $6$, $8$, $10$, $16$\} \cite{bunge2013texture}. For comparison, we show the corresponding Sinkhorn distance.

The value ranges of the different analyzed distances differ due to different representations. To compensate this effect the distances are normalized to the value of the distance at step 19 (maximum distance value): 
\begin{equation}
\mathscr{D} = \mathscr{D}(f^{(0)}(g), f^{(k)}(g)) / \mathscr{D}(f^{(0)}(g), f^{(19)}(g))
\end{equation}

\begin{figure} [H]
	\centering
	\includegraphics[width=0.99\textwidth]{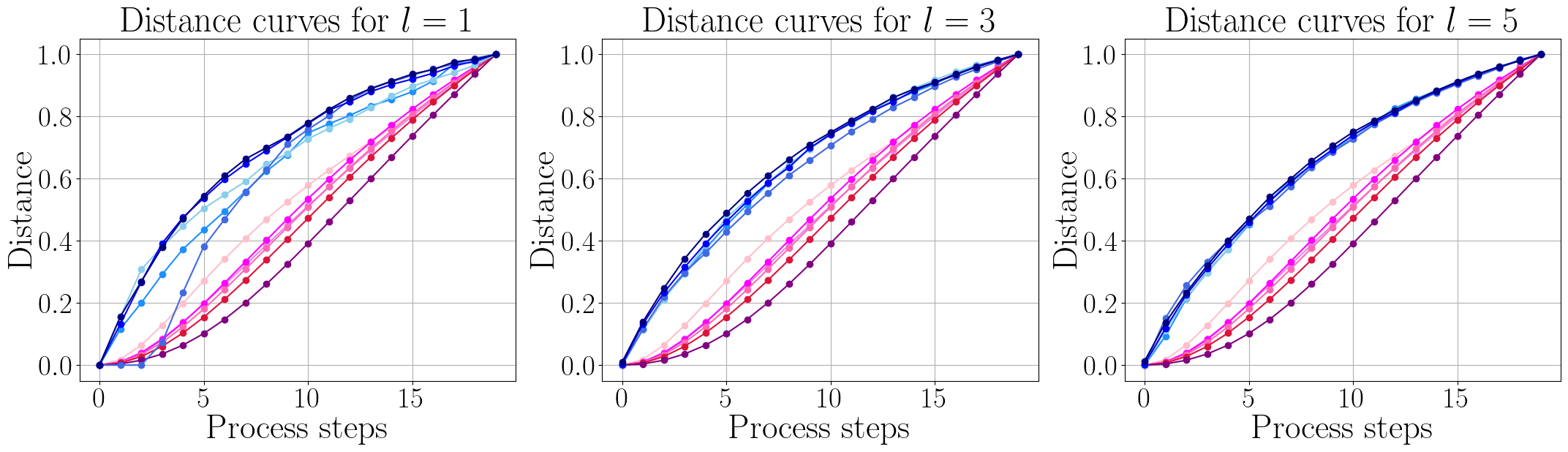}
	\includegraphics[width=0.90\textwidth]{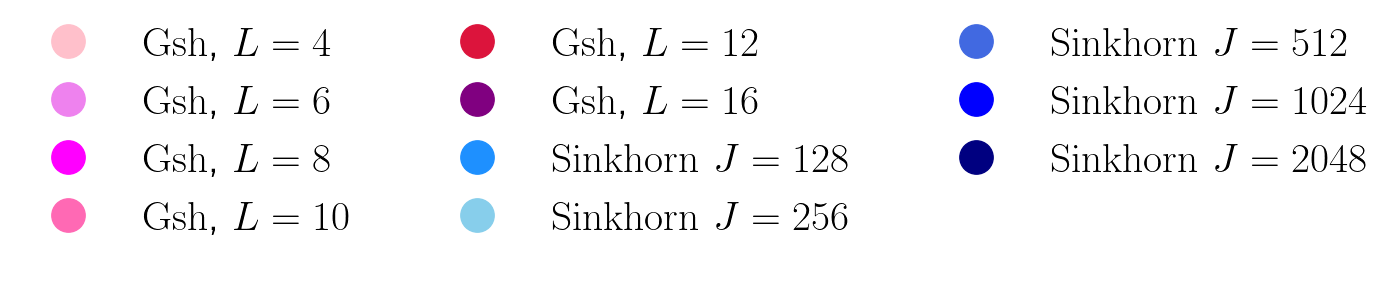}
	\caption{Distances over process step curves for the textures with $512$ orientations ($n_{\mathrm{oris}}=512$). In each of the charts (a), (b) and (c) the GSH distance (red) is plotted for L $\in$ \{$4$, $6$, $8$, $10$, $16$\}. The Sinkhorn distance curves (blue) for number of bins $J \in$ \{$128$, $256$, $512$, $1024$, $2048$\} are shown in (a) for the soft-assignment $l=1$, in (b) for $l = 3$ and in (c) for $l = 5$.}
  \label{img:gsh-distances}
\end{figure}

The results for the GSH distance are shown over process step numbers in Fig. \ref{img:gsh-distances}. It can be seen that the Sinkhorn distance increases stronger in the first steps and then tends to saturate at later steps. In contrast, the GSH distance shows a small increase at the beginning and increases stronger after certain process steps, resulting in an approximately linear increase from step 8 onwards. In general, the GSH distance is strictly monotonic and smooth. The curves show that the GSH distance measure is less sensitive to small textures distances than the Sinkhorn distance measure, while the latter tends to saturate at higher distances. 

\subsection{Learning texture-property relation}
\label{ss:Results-Spm}

The results of the texture-property relations are summarized in Tab. \ref{tab:summary-of-spm}. The coefficient of determination $R^2$ is used for comparison, as well as the regression errors MAE$_\mathrm{E}$ and MAE$_{\tilde{\mathrm{R}}}$, which denote the mean absolute error (MAE) between the true and predicted Young's moduli and $\tilde{R}$-values. An MAE of Young's moduli $\leq 1000$ and of $\tilde{R}$ of $\leq 0.1$ is sufficient in most applications. Additionally, the numbers of input features of the used texture representations are noted. It is shown that texture-property relations with a sufficient prediction quality can be achieved for all texture representations. By considering the orientation histograms, better results are obtained with the soft-assignment $l=3$. Comparing all texture representations, the estimator quality for GSH for $L=16$ is best considering $R^2$ while pole figures achieves the best result considering  MAE. The distribution of the property values of the learning data set is shown in Tab. \ref{tab:property-ranges}.

\begin{table}[H]
	\caption{Results of the investigated textures for learning the texture-property relations. Regression error MAE$_\mathrm{E}$ is given in $[$GPa$]$, Regression error MAE$_{\tilde{\mathrm{R}}}$ in $[$-$]$ and Regression score $R^2$ in $[$\%$]$. The best results for the metrics are noted in bold.}
	\begin{center}
	\begin{tabular}{ l l c c c c}
		\toprule
		Texture	& Number of features	& $R^2$	&	MAE$_\mathrm{E}$	& MAE$_{\tilde{\mathrm{R}}}$	\\ \midrule
		\makecell[l]{Orientation histograms \\ for $J=512, l=1$}	& 512	& 98.2	& 0.329	& 0.022		\\ 
		\makecell[l]{Orientation histograms \\ for $J=512, l=3$}	& 512	& 99.5	& 0.125	& 0.012		\\ \midrule 
		GSH for $L=10$ 			     								& 80	& 99.3	& 0.152	& 0.015		\\ 
		GSH for $L=16$ 			     								& 280	& \textbf{99.7}	& 0.087	& \textbf{0.011}	\\ \midrule
		Pole figures   											& $64 \times 64 \times 3$  	& 99.6	& \textbf{0.080}& \textbf{0.011}	\\
		\bottomrule
		\end{tabular}							
	\end{center}				
	\label{tab:summary-of-spm}
\end{table}		

\begin{table}[H]
	\caption{Distribution of the property space for the underlying data set. The E-Values are given in $[$GPa$]$ and the $\tilde{R}$-Values are given in $[$-$]$.}
	\begin{center}
		\begin{tabular}{ c c c c c c}
		\toprule
			Property 			& Min. Value 	& Max. Value 	& Range	& Mean	& Std. 	\\ \midrule
			E$_{11}$			& 204			& 231		 	& 27	& 216	& 3.14	\\ 
			E$_{22}$			& 206			& 234		 	& 28	& 218	& 3.42	\\ 
			E$_{33}$			& 204			& 229		 	& 25	& 217	& 3.27	\\ \midrule 
			$\tilde{R}_{11}$	& 0.373			& 4.024		 	& 3.651	& 1.084	& 0.284	\\ 
			$\tilde{R}_{22}$	& 0.342			& 2.882		 	& 2.540	& 0.964	& 0.167	\\ 
			$\tilde{R}_{33}$	& 0.239			& 2.210		 	& 1.971	& 0.931	& 0.194	\\ 
		\bottomrule
		\end{tabular}
	\end{center}
	\label{tab:property-ranges}
\end{table}

\section{Discussion}
\label{s:Discussion}
The findings of this work will be discussed under three aspects: (1) ODF-based texture representations, (2) texture distances measured with representations as discussed in (1), and (3) effects of texture representations on learning of texture-property relations.    

\paragraph{ODF-based texture representations:} Pole figures are the richest representation of the investigated textures in this study, as they are represented as image data. In this representation, the neighborhood relations of the orientations are represented by the neighboring pixels in the image. Moreover, the visualization of the ODF by pole figures is interpretable by domain experts, which in turn allows them to interpret texture-property relations represented by neural regression networks.In contrast, neighborhood relations are not reflected by orientation histograms, since the order of bins does not reflect the order of orientations, cf. Section \ref{sss:ohg-texture-description}. Orientation relationships are only represented by orientation histograms when the orientations of the bin centers are included. The representations of ODFs by GSH coefficients does not contain any information about orientation relationships. When using GSHs, neighborhood relations between orientations are implicitly represented by the base functions, which belong to the GSH coefficients. When expanding an ODF using GSH with only low order coefficients, sharp textures can not be represented very well. Besides pole figure images, plots showing sections through the Euler space are also often used to visualize the orientation distribution. However, due to the distortion of the Euler space, this representations is disadvantageous for representing and processing crystallographic texture data.

\paragraph{Texture distance measures:} Distances between textures using the \textit{distance measurement evaluation set} described in Section \ref{ss:Texture-data-generation} are evaluated by means of a stepwise process, which pushes the texture further apart from the initial texture at each step. When constantly applying tensile load increments on a material with initial texture $f^{(0)}(g)$ as it is done in the example process (cf. Section \ref{sec:materialmodel}), the distance between texture $f^{(0)}(g)$ and $f^{(k)}(g)$ is expected to be smoothly and monotonically increasing with step number $k$. The GSH-based distance measure as well as the distance measures based on pole figures show this behavior. The monotonic requirement of the distance measure is also preserved by the Sinkhorn distance of the histogram representation, while it is violated by the Chi-Squared distance measure at soft-assignment $l=1$ of the histogram representation. The Chi-Squared distance measure is often not smooth (see Fig. \ref{img:n-base-512}), independent of the soft-assignment parameter, while the Sinkhorn distance is smooth in all of the presented studies (except for the single crystal example shown in Fig. \ref{img:n-base-512} (g)). 

Another disadvantage of the Chi-Squared distance is its high sensitivity with respect to a shift of orientations from one bin to another one. This is less pronounced for dense histograms (i.e. where there is a non-zero number in each bin) than for sparse histograms. This is best seen by looking at extreme cases of three sparse histograms: Histogram $f_{\boldsymbol{\mathrm{b}}}^{(1)}(g)$ has orientations only in one single bin $o_1$, otherwise all entries are zero. Histogram $f_{\boldsymbol{\mathrm{b}}}^{(2)}(g)$ has orientations only in an immediate neighbor bin of $o_1$. Histogram $f_{\boldsymbol{\mathrm{b}}}^{(3)}(g)$ has orientations only in any other bin. The Chi-Squared distances in both cases $\mathscr{D}_{\mathrm{\chi^2}}(f_{\boldsymbol{\mathrm{b}}}^{(1)} (g), f_{\boldsymbol{\mathrm{b}}}^{(2)} (g)) = \mathscr{D}_{\mathrm{\chi^2}}(f_{\boldsymbol{\mathrm{b}}}^{(1)} (g), f_{\boldsymbol{\mathrm{b}}}^{(3)} (g)) = 2$ and thus maximal (following Eq. (\ref{eq:chi2_whd})), no matter how close the unit quaternions of the bins are. 

In general, it can be seen that the soft-assignment parameter has a strong effect on the Chi-Squared distance measure, as it incorporates local orientation distances by distributing histogram weights to several bins. This is artificially countering the disadvantage of sensibility against
the shift of orientations. The Sinkhorn distance, in contrast, is not affected much by the soft-assignment parameter as it incorporates also the distances between the orientations of the bin representatives (cf. Eq(\ref{eq:quat-distance-emd})). This reflects the neighborhood relations of the orientation space over which the ODF is spanned. Nevertheless, larger values of soft-assignment $l$ yield smoother distance curves. In the case of Chi-Squared distance the usage of $l>1$ is mandatory for the above mentioned reasons.

In addition, the number of histogram bins of the orientation histogram has a significant effect on the Chi-Squared distance measure, which can be seen in Fig. \ref{img:n-base-512} (a) - (c)  and Fig. \ref{img:n-base-512} (d) - (e): the higher the number of histogram bins, the lower the distance is. For higher number of histogram bins (e.g. $J=2048$), the Chi-Squared distance even tends to converge towards the Sinkhorn distance. The Sinkhorn distance in-turn is much less sensitive to the number of histogram bins. 

Regarding the single crystal texture study (Initial Texture 3 ($n_{\mathrm{oris}}$=$1$)), the orientation histogram-based distance measures do not show a smooth behavior, because the crystal rotation after loading cannot be tracked adequately by the orientation histograms. Therefore, the Chi-Squared and the Sinkhorn distance jumps at certain load steps, independently of the number of histogram bins and soft-assignment parameter. However, because of the bin-wise comparison, the Chi-Squared distance reaches quickly it's maximum value. In contrast, the Sinkhorn distance does not reach the maximum value at all and converges almost to the same value for different number of histogram bins and soft-assignments. Moreover, the Sinkhorn distance is closer to the quaternion distance, which can be considered to be the ground truth in this case. 

In contrast to the orientation histogram-based distance measures, the pole figure distance shows a rather smooth behavior over the loading steps, as the rotation of the single crystal is tracked by the pole figure. This is due to the generation of pole figures by using a Gaussian kernel which spreads the orientations over several pixels. The distance curves resulting from pole figures are monotonic and smooth. Also the GSH distance shows a smooth behavior in all cases, as it is based on a smooth function approximation (cf. Eq. (\ref{eq:gsh-5})). However, in Fig. \ref{img:gsh-distances}, it can be seen that the GSH distance behaves contrary to the Sinkhorn distance, as it has a slight increase at the beginning of the loading steps and a rather strong increase later. This characteristic cannot be influenced by the degree $L$ of the GSH distance and does not seem to have a strong effect in this example. It can be seen from Fig. \ref{img:gsh-distances} that the GSH distance measure has lower resolution power at small texture distance than the Sinkhorn distance measure while it is higher at larger texture distances than Sinkhorn.

\paragraph{Estimation of texture-property relations:} Besides evaluating different texture representations for texture distance measurement their suitability for modeling texture-property relations was investigated in Section \ref{ss:Structure-Property-Mapping}. The results show that sufficient prediction quality can be achieved for all texture representations (see Table \ref{tab:summary-of-spm}), whereas the best results are obtained with the textures represented as pole figures (considering MAE) and GSH with $L=16$ (considering $R^2$).

The pole figure representation allows the neural network models to exploit the pixel-represented neighborhood relationships within the textures through convolutional operations, which explains the superior prediction quality. However, the pole figure image data must be transformed by a backbone to a vectorized latent space by means of convolutional and pooling layers, before the neural network head can perform the regression to the property values. The additional backbone layers make processing of pole figures more complex than using histograms as input, which results in higher model training efforts. 

In contrast, the orientation histograms can be directly used as input for the neural regression network, reducing the overall network complexity. Investigations were performed for the number of histogram bins $J=512$ and soft-assignment $l=1$ and $l=3$, where the better result is achieved with $l=3$. The improvement with $l=3$ can be explained by the approximate consideration of neighborhood relationships by higher order smoothing via soft assignment. The number of bins of $J=512$ was chosen as compromise between computational cost and representation quality. 

The GSH representations can be directly used as input as well. The investigation was performed using GSH representations of the degrees $L=16$ and $L=10$, which comprise of $280$ and $80$ features respectively. A better prediction quality is achieved with $L=16$, but with $L=10$ sufficient results are already achieved with its lower number of features, requiring a less complex neural network model. Better prediction results than with orientation histograms can be achieved by GSH-based models with lower number of features due to the implicit representation of orientation neighborhood relations in the base functions of the GSHs.

\section{Summary and outlook}
\label{s:Summary}
The measurement of texture distances is essential for any task, where knowledge about the assessment of texture similarities is required. Examples are the design of processes, which are supposed to reach a given a target texture, and modeling of texture-property relations for materials design aiming to optimize textures for given properties.

In the literature on material science and engineering, only a few studies can be found related to distance measures between textures. Therefore this has been investigated in more detail in this paper for crystallographic textures, which are commonly described by the orientation distribution function (ODF). Methods of distance measurement based on representations of the orientation distribution function such as generalized spherical harmonics, orientation histograms and pole figures have been developed and investigated: (1) explicit distance measure between two ODFs expanded in GSH, (2) Sinkhorn distance measure and (3) Chi-Squared distance measure for ODFs represented by orientation histograms and (4) distance measure for ODFs represented as pole figures.

By evaluating the texture distances resulting from an example process, we were able to figure out the advantages and drawbacks using the discussed texture representations and distance measures. The process generates textures with increasing distance from the initial texture at each process step. This produces a monotonically increasing distance-over-process-step-number curve. By investigating the corresponding distance curves calculated from different texture representations and measures, it turns out that using histogram-based texture representations with the Sinkhorn distance are the best compromise regarding precision and computational complexity. It has been shown that the Sinkhorn distance curves are close to the ground truth quaternion distance curve for single crystals. Comparing the Sinkhorn distance and the GSH distance, the Sinkhorn distance has higher sensitivity with small distances (at the beginning of the process curve) and lower sensitivity with large distances (at the end of the process curve), while the behavior of the GSH distance is opposite (cf. Section \ref{sss:Texture-Distance-Measurement-GSH} and Fig. \ref{img:gsh-distances}). For this reason, the Sinkhorn distance should be chosen when a high resolution of small distances is more important, and the GSH distance when large distances have to be better resolved. The comparison of the pole figures with the histogram based textures shows that the pole figure distance achieves better results than the Chi-Squared distance but not as good as the Sinkhorn distance. When the additional visualization of texture representation is desired, pole figures can be used.

Considering the suitability of the investigated texture representations for learning texture-property relations, using GSHs achieves similar quality than using pole figures, but with a lower computing complexity. Both are superior to using orientation histograms which make only limited use of orientation neighborhoods. 

In the present work, we applied our methods on data from mean-field simulations. The next step is to apply distance measurements as well as learning the structure-property relationship on spatially resolved data from full-field simulations. For the latter, the problem arises that the computational efforts of simulation and of machine learning are much higher and therefore fewer data can be generated (via full-field simulations) and processed. Due to the findings of this work, the Chi-Sqaured distance will be replaced by the Sinkhorn distance in the loss function of the SMTL model \cite{iraki2021multi} in our future work.

\section*{Data availability}
The texture-property data set used for learning texture-property relations is made available publicly via Fraunhofer Fordatis repository at \url{https://fordatis.fraunhofer.de/handle/fordatis/319} \cite{processdata2023}.

\section*{Acknowledgements}
We would like to thank the German Research Foundation (DFG) for funding the presented work, which was carried out within the research project number 415804944: 'Taylored Material Properties via Microstructure Optimization: Machine Learning for Modelling and Inversion of Structure-Property-Relationships and the Application to Sheet Metals'. Also we would like to thank Prof. Surya Kaldindi for his support to derive the GSH distance.

\bibliographystyle{ieeetr}
\bibliography{paper}   

\clearpage
\appendix
\section{GSH Distance}
\label{sec:Appendix-Gsh-Distance}

The distance between two ODFs $f^{(1)} (g)$ and $f^{(2)} (g)$ can be expressed according to Eq. (\ref{eq:global_comparison}) as

\begin{equation}
\begin{aligned}
	\mathscr{D}_{\mathrm{odf}}(f^{(1)} (g),f^{(2)} (g)) & = \frac{1}{8 \pi^2} \int\limits_{0}^{2\pi} \int\limits_{0}^{\pi} \int\limits_{0}^{2\pi} \\
	& \Bigl(f^{(1)}(\varphi_1, \Phi, \varphi_2) - f^{(2)}(\varphi_1, \Phi, \varphi_2)\Bigr)^2 \\
	& d \varphi_1 \sin(\Phi) d \Phi d \varphi_2.
\end{aligned}
\label{eq:gsh-app-1}
\end{equation}

When the densities $f^{(1)} (g)$ and $f^{(2)} (g)$ are expanded in two series of GSH base functions

\begin{equation}
f_{\mathrm{gsh}}^{(1)}(g) = \sum_{l=0}^{\infty} \sum_{m=-l}^{+l} \sum_{n=-l}^{+l} C_{l}^{mn} e^{im\varphi_2} P_{l}^{mn} (cos(\Phi)) e^{in\varphi_1}
\label{eq:gsh-app-2}
\end{equation}

and

\begin{equation}
f_{\mathrm{gsh}}^{(2)}(g) = \sum_{l=0}^{\infty} \sum_{m=-l}^{+l} \sum_{n=-l}^{+l} \tilde{C}_{l}^{mn} e^{im\varphi_2} P_{l}^{mn} (cos(\Phi)) e^{in\varphi_1}.
\label{eq:gsh-app-3}
\end{equation}

The distance measure of Eq. (\ref{eq:gsh-app-1}) becomes
\begin{equation}
\begin{aligned}
\mathscr{D}_{\mathrm{gsh}}(f_{\mathrm{gsh}}^{(1)}(g),f_{\mathrm{gsh}}^{(2)}(g)) & = \frac{1}{8 \pi^2}  \int\limits_{0}^{2\pi} \int\limits_{0}^{\pi} \int\limits_{0}^{2\pi}\\
         & \Biggl( \sum_{l=0}^{\infty} \sum_{m=-l}^{+l} \sum_{n=-l}^{+l} (C_{l}^{mn} - \tilde{C}_{l}^{mn}) e^{im\varphi_2} P_{l}^{mn} (\cos(\Phi)) e^{in\varphi_1} \Biggr)^2\\
         &  d \varphi_1 \sin(\Phi) d \Phi d \varphi_2.
\end{aligned}
\label{eq:gsh-app-4}
\end{equation}

Substituting $u = cos(\Phi)$ yields $du = -sin(\Phi) d \Phi$ and 
\begin{equation}
\begin{aligned}
\mathscr{D}_{\mathrm{gsh}}(f_{\mathrm{gsh}}^{(1)}(g),f_{\mathrm{gsh}}^{(2)}(g)) & = \frac{1}{8 \pi^2}  \int\limits_{0}^{2\pi} \int\limits_{-1}^{+1} \int\limits_{0}^{2\pi} \\
		& \Biggl( \sum_{l=0}^{\infty} \sum_{m=-l}^{+l} \sum_{n=-l}^{+l} (C_{l}^{mn} - \tilde{C}_{l}^{mn}) e^{im\varphi_2} P_{l}^{mn} (u) e^{in\varphi_1} \Biggr)^2 \\
		& d \varphi_1 du d \varphi_2.
\end{aligned}
\label{eq:gsh-app-5}
\end{equation}


All mixed terms contain expressions of
\begin{equation}
\begin{split}
\int\limits_{0}^{2\pi} \int\limits_{-1}^{+1} \int\limits_{0}^{2\pi} e^{i(m+v)\varphi_2} P_{l}^{mn} (u) P_{p}^{vw} (u) e^{i(n+w)\varphi_1} d \varphi_1 du d \varphi_2 \quad \\
\textrm{with} \quad m \neq v, n \neq u, l \neq w \\
= \int\limits_{0}^{2\pi} e^{i(m+v)\varphi_2} d \varphi_2 \int\limits_{0}^{2\pi} e^{i(n+w)\varphi_1} d \varphi_1 \int\limits_{-1}^{+1} P_{l}^{mn} (u) P_{p}^{vw} (u) du = 0 \quad \\
\textrm{for} \quad m \neq v, n \neq w, l \neq p.
\end{split}
\label{eq:gsh-app-6}
\end{equation}

The mixed terms therefore all vanish due to orthogonality and only the squared terms remain
\begin{equation}
\begin{aligned}
\mathscr{D}_{\mathrm{gsh}}(f_{\mathrm{gsh}}^{(1)}(g),f_{\mathrm{gsh}}^{(2)}(g)) & = \frac{1}{8 \pi^2} \int\limits_{0}^{2\pi} \int\limits_{-1}^{+1} \int\limits_{0}^{2\pi} \\
	&  \Biggl[ \sum_{l=0}^{\infty} \sum_{m=-l}^{+l} \sum_{n=-l}^{+l} (C_{l}^{mn} - \tilde{C}_{l}^{mn})^2 (e^{im\varphi_2})^2 (P_{l}^{mn} (u))^2 (e^{in\varphi_1})^2 \Biggr] \\ 
	& d \varphi_1 du d \varphi_2.
\end{aligned}
\label{eq:gsh-app-7}
\end{equation}

The integration of the terms containing the integrands gives
\begin{equation}
\begin{split}
\int\limits_{0}^{2\pi} \int\limits_{-1}^{+1} \int\limits_{0}^{2\pi} (e^{im\varphi_2})^2 (P_{l}^{mn} (u))^2 (e^{in\varphi_1})^2 d \varphi_1 d u \varphi_2 \\
= \int\limits_{0}^{2\pi} (e^{im\varphi_2})^2 d \varphi_2 \int\limits_{0}^{2\pi} (e^{in\varphi_1})^2 d \varphi_1 \int\limits_{-1}^{+1} (P_{l}^{mn} (u))^2 du
\end{split} 
\label{eq:gsh-app-8}
\end{equation}

The integrads $(e^{im\varphi_2})^2$ and $(e^{in\varphi_1})^2$ are always equal to $1$. 



Therefore
\begin{equation}
\int\limits_{0}^{2\pi} (e^{im\varphi_2})^2 d \varphi_2 \int\limits_{0}^{2\pi} (e^{in\varphi_1})^2 d \varphi_1 = 4 \pi^2 
\label{eq:gsh-app-9}
\end{equation}

The integral over the Legendre polynomials is
\begin{equation}
\int\limits_{-1}^{+1} (P_{l}^{mn} (u))^2 du = \frac{2}{2l+1}
\label{eq:gsh-app-10}
\end{equation}

Inserting all terms into
\begin{equation}
\begin{aligned}
\mathscr{D}_{\mathrm{gsh}}(f_{\mathrm{gsh}}^{(1)}(g),f_{\mathrm{gsh}}^{(2)}(g)) & =  \frac{1}{8 \pi^2}  \int\limits_{0}^{2\pi} \int\limits_{-1}^{+1} \int\limits_{0}^{2\pi} \\
	& \Biggl( \sum_{l=0}^{\infty} \sum_{m=-l}^{+l} \sum_{n=-l}^{+l} (C_{l}^{mn} - \tilde{C}_{l}^{mn})^2 (e^{im\varphi_2})^2 (P_{l}^{mn} (u))^2 (e^{in\varphi_1})^2 \Biggr) \\
	& d \varphi_1 du d \varphi_2.
\end{aligned}
\label{eq:gsh-app-11}
\end{equation}

The distance measure between orientation densities $f_{\mathrm{gsh}}^{(1)}(g)$ and $f_{\mathrm{gsh}}^{(2)}(g)$, when expressed as series expansion of generalized spherical harmonics is then finally
\begin{equation} 
\mathscr{D}_{\mathrm{gsh}}(f_{\mathrm{gsh}}^{(1)}(g),f_{\mathrm{gsh}}^{(2)}(g))  =  \sum_{l=0}^{\infty} \sum_{m=-l}^{+l} \sum_{n=-l}^{+l} \frac{1}{2l + 1} (C_{l}^{mn} - \tilde{C}_{l}^{mn})^2.
\label{eq:gsh-app-12}
\end{equation}

\section{Taylor-type material model}
\label{sec:Appendix-Material-Model}
The basic form of the Taylor-type material model used in this study originates from the description in \cite{Kalidindi.1992}. Further descriptions of the used material model can be found in \cite{dornheim2020structureguided} and \cite{iraki2021multi} and are as follows. The operators $\cdot$ and $:$ denote dot and the double dot product.

For $n$ crystals, the volume averaged stress is defined by
\begin{equation}
\boldsymbol{\overline{T}} = \frac{1}{V}\sum_{i=1}^n \boldsymbol{T}^{(i)}V^{(i)},
\end{equation}
with $\boldsymbol{T}^{(i)}$ being the Cauchy stress tensor of a single crystal $i$ with volume $V^{(i)}$. The Cauchy stress tensor $\boldsymbol{T}$ is derived from the stress tensor in the intermediate configuration using a multiplicative decomposition of the deformation gradient in its elastic and plastic part ($\boldsymbol{F}=\boldsymbol{F}_\mathrm{e}\cdot \boldsymbol{F}_\mathrm{p}$):
\begin{equation}
\boldsymbol{T}^*=\frac{1}{2} ~ \mathbb{C}:(\boldsymbol{F}_\text{e}^T\cdot\boldsymbol{F}_\text{e}-\boldsymbol{I}).
\label{eq:elasticity}
\end{equation}
Here, $\boldsymbol{I}$ denotes the second order identity tensor and $\mathbb{C}$ denotes the forth order elastic stiffness tensor. When applying to cubic crystal symmetry, $\mathbb{C}$ is composed of three independent parameters, which are $C_{11}$, $C_{12}$, and $C_{44}$. To transform $\boldsymbol{T}$ into $\boldsymbol{T^*}$, the following relation is used:
\begin{equation}
\boldsymbol{T^*} = \boldsymbol{F}_\text{e}^{-1}\cdot(\operatorname{det}(\boldsymbol{F}_\text{e})~\boldsymbol{T})\cdot\boldsymbol{F}_\text{e}^{-\top}.
\end{equation}

The plastic velocity gradient $\boldsymbol{L}_\mathrm{p}$, which is used to described the evolution of the plastic deformation, is defined as the sum of the shear rates $\dot{\gamma}^{(\eta)}$ on the slip systems $\eta$ \cite{Rice.1971}
\begin{equation}
\boldsymbol{L}_\text{p} = \dot{\boldsymbol{F}}_\text{p} \cdot\boldsymbol{F}_\text{p}^{-1}=\sum_\eta \dot{\gamma}^{(\eta)} \boldsymbol{m}^{(\eta)} \otimes \boldsymbol{n}^{(\eta)}.
\label{eq:plastic_velocity_gradient}
\end{equation}
The slip systems are defined by the slip plane normal $\boldsymbol{n}^{(\eta)}$ and the slip direction $\boldsymbol{m}^{(\eta)}$. For the purpose of this work, we used the slip system families for body-centered cubic crystals ($\{110\}\langle111\rangle$ and $\{112\}\langle111\rangle$ in Miller index notation).

A phenomenological power law, as is for example described in \cite{Asaro.1985}, is used to determine the slip rates: 
\begin{equation}
\dot{\gamma}^{(\eta)}=\dot{\gamma}_0 \left| \frac{\tau^{(\eta)}}{r^{(\eta)}} \right|^{1/m}\text{sign}(\tau^{(\eta)}),
\end{equation}
where $r^{(\eta)}$, $\dot{\gamma}_0$, and $m$ are material dependent parameters for the slip resistance, reference shear rate, and the shear rate sensitivity, respectively. Furthermore, the resolved shear stress, denoted by $\tau^{(\eta)}$, is defined by Schmid's law
\begin{equation}
\tau^{(\eta)}=((\boldsymbol{F}_\text{e}^T\cdot\boldsymbol{F}_\text{e})\cdot\boldsymbol{T}^*):(\boldsymbol{m}^{(\eta)}\otimes\boldsymbol{n}^{(\eta)}).
\end{equation}

The evolution of slip resistance is described by
\begin{equation}
\dot{r}^{(\eta)} = \frac{\mathrm{d}\hat{\tau}^{(\eta)}}{\mathrm{d}\Gamma}\sum_\zeta \hat{q}_{\eta\zeta}|\dot{\gamma}^{(\zeta)}|,
\end{equation}
with an interaction matrix $\hat{q}_{\eta\zeta}$ taking into account self and latent hardening. This matrix is composed of diagonal elements equal to 1.0 and off-diagonal elements $q_1$ and $q_2$ (cf. \cite{baiker2014determination}). In addition, $\Gamma$ denotes the accumulated plastic shear, which is defined by
\begin{equation}
\Gamma = \int_0^t \sum_\eta \left| \dot{\gamma}^{(\eta)} \right| \mathrm{d}t.
\end{equation}

Finally, an extended Voce-type model is used to describe the hardening behavior \cite{Tome.1984}:
\begin{equation}
\hat{\tau}^{(\eta)}=\tau_0+(\tau_1+\vartheta_1\Gamma)(1-e^{-\Gamma\vartheta_0/\tau_1}),
\label{eq:voce-type_hardening}
\end{equation}
as is used for example in \cite{pagenkopf2016virtual} and \cite{baiker2014determination}. The used model contains four material dependent parameters $\tau_0$, $\tau_1$, $\vartheta_0$, and $\vartheta_1$, which are calibrated to DC04 steel.
The material dependent parameters that are used in this study are defined as in \cite{dornheim2020structureguided}. These are summarized in Table \ref{tab:mat_parameters}.

\begin{table}
	\centering
	\caption{Material dependent parameters used in this study (cf. \cite{dornheim2020structureguided})}
	\begin{tabular}{|c| c c c c c c c c c c c |}
	\hline
	parameter & $C_{11}$ & $C_{12}$ & $C_{44}$ & $\dot{\gamma}_0$ & $m$ & $\tau_0$ & $\tau_1$ & $\vartheta_0$ & $\vartheta_1$ & $q_1$ & $q_2$ \\
	\hline
	value & 226 & 140 & 116 & 0.001 & 0.02 & 90 & 32 & 250 & 60 & 1.4 & 1.4 \\ 
	\hline
	unit & GPa & GPa & GPa & s$^{-1}$ & - & MPa & MPa & MPa & MPa & - & - \\
	\hline
	\end{tabular}
	\label{tab:mat_parameters}
\end{table}

The reorientation of the crystals is calculated based on a rigid body rotation $\boldsymbol{R}_\mathrm{e}$ \cite{ling2005numerical}:
\begin{equation}
\boldsymbol{R}_1 = \boldsymbol{R}_\mathrm{e} \boldsymbol{R}_0,
\end{equation}
with the rotation matrices $\boldsymbol{R}_0$ describing the original crystal orientation and $\boldsymbol{R}_1$ describing the orientation after deformation. $\boldsymbol{R}_\mathrm{e}$ is derived using the polar decomposition of $\boldsymbol{F}_\mathrm{e}$:
\begin{equation}
\boldsymbol{R}_\mathrm{e} = \boldsymbol{F}_\mathrm{e} \cdot \boldsymbol{U}_\mathrm{e}^{-1}
\end{equation}
with the elastic part of the right Cauchy-Green tensor $\boldsymbol{U}_\mathrm{e}$. The orientation of a crystal is therefore expected to change smoothly for an applied load. For tensile load increments, we furthermore expect the distances between the initial texture and the evolving texture to increase monotonically.

\section{SPM-Details}
\label{sec:SPM-Details}

\begin{table}
	\centering
	\caption{Topologies of the developed neural networks models for learning the structure-property relations.}
	\begin{tabular}{| l | l | l | l |}	
		\hline	
		Texture	& Layer	& Layer type 		& Output dim.	\\
		\hline	
		\hline	
		\multirow{6}{*}{(a) Orientation histogram} 	
				& 0		& Input layer		& $512 \times 1$ 	\\
				& 1 	& Fc + Tanh 		& $256 \times 1$	\\
				& 2 	& Fc + Tanh			& $128 \times 1$	\\
				& 3 	& Fc + Tanh			& $64 \times 1$	\\
				& 4		& Fc + Tanh			& $20 \times 1$	\\
				& 5 	& Fc + Tanh 		& $10 \times 1$	\\
				& 6 	& Fc 		 		& $6 \times 1$	\\				
		\hline	

		\multirow{4}{*}{(b) GSH, L=10} 	
				& 0 	& Input layer	 	& $80 \times 1$ 	\\
				& 1 	& Fc + Tanh 		& $40 \times 1$	\\
				& 2		& Fc + Tanh			& $20 \times 1$	\\
				& 3		& Fc + Tanh			& $10 \times 1$	\\
				& 4 	& Fc 				& $6 \times 1$	\\				
		\hline	

		\multirow{6}{*}{(c) GSH, L=16} 	
				& 0 	& Input layer	 	& $280 \times 1$ 	\\
				& 1 	& Fc + Tanh 		& $160 \times 1$	\\
				& 2 	& Fc + Tanh			& $80 \times 1$	\\
				& 3 	& Fc + Tanh			& $40 \times 1$	\\
				& 4		& Fc + Tanh			& $20 \times 1$	\\
				& 5		& Fc + Tanh			& $10 \times 1$	\\
				& 6		& Fc 				& $6 \times 1$	\\				
		\hline	

		\multirow{10}{*}{(d) Pole figures} 
			& 0 		& Input layer 		& $64 \times 64 \times 3$		\\
			& 1 		& 2dConv + Tanh		& $64 \times 64 \times 6$		\\
			& 2 		& MaxPool2d			& $32 \times 32 \times 6$ 	\\
			& 3 		& 2dConv + Tanh		& $32 \times 32 \times 12$	\\
			& 4 		& MaxPool2d			& $16 \times 16 \times 12$	\\
			& 5 		& Vectorized		& $3072 \times 1$			\\
			& 6 		& Fc + Tanh			& $200 \times 1$ 			\\
			& 7 		& Fc + Tanh			& $100 \times 1$			\\
			& 8 		& Fc + Tanh			& $20 \times 1$			\\
			& 9 		& Fc + Tanh			& $10 \times 1$			\\
			& 10 		& Fc 				& $6 \times 1$				\\			
		\hline
					
	\end{tabular}
	\label{tab:Spm-Topologies}
\end{table}

\begin{figure}
  \centering
  \subfloat[][]{\includegraphics[width=0.66\textwidth]{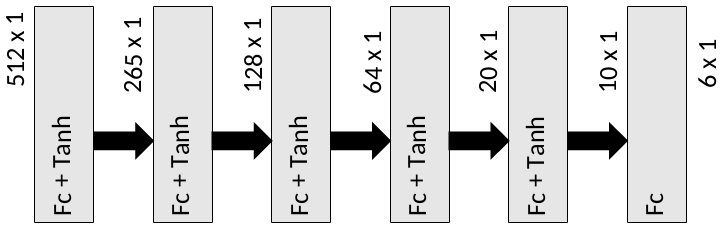}}
  \qquad
  \subfloat[][]{\includegraphics[width=0.45\textwidth]{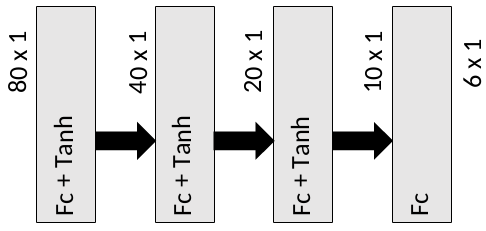}}
  \qquad
  \subfloat[][]{\includegraphics[width=0.66\textwidth]{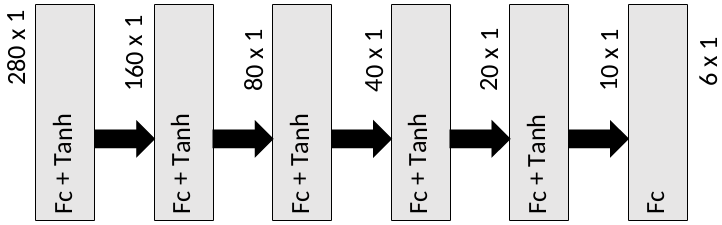}}
  \qquad
  \subfloat[][]{\includegraphics[width=0.85\textwidth]{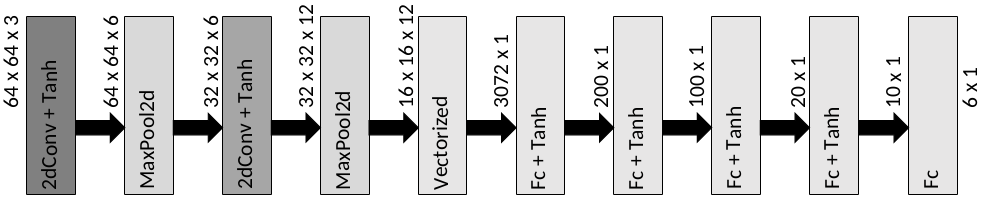}}
  \caption{Topologies of the neural networks for learning the texture-property relations by the texture representations (a) Orientation histograms, (b) GSH for $L$=10, (c) GSH for $L$=16 and (d) Pole figures.}
  \label{img:Spm-Topologies}  
\end{figure}

\end{document}